\documentclass{aa}

\usepackage{graphicx}
\usepackage{txfonts}
\usepackage{natbib}     
\usepackage{xcolor} 
\usepackage{threeparttable} 
\usepackage{comment} 

\bibpunct{(}{)}{;}{a}{}{,} 

\begin{document}

\title{Detection of Fe and evidence for TiO in the dayside emission spectrum of WASP-33b}

\author{D. Cont\inst{1}
	\and
	F. Yan\inst{1}
	\and
	A. Reiners\inst{1}
	\and
	N. Casasayas-Barris\inst{2,3,4}
	\and
	P.~Molli\`ere\inst{5}
	\and   		
	E. Pall\'e\inst{2,3}
	\and
	Th. Henning\inst{5}
	\and
	L.~Nortmann\inst{1}
	\and
	M.~Stangret\inst{2,3}
	\and
	S.~Czesla\inst{6}
	\and
	M.~L\'opez-Puertas\inst{7}
	\and  
	A.~S\'anchez-L\'opez\inst{4}
	\and
	F.~Rodler\inst{8}
	\and  
	I.~Ribas\inst{9,10}
	\and
	A.~Quirrenbach\inst{11}
	\and
	J.~A.~Caballero\inst{12}
	\and
	P.~J.~Amado\inst{7}
	\and
	L.~Carone\inst{5}
	\and
	J.~Khaimova\inst{1}
	\and
	L.~Kreidberg\inst{5}
	\and
	K.~Molaverdikhani\inst{11,5}
	\and
	D.~Montes\inst{13}
	\and
	G.~Morello\inst{2}
	\and
	E.~Nagel\inst{6,14}
	\and
	M.~Oshagh\inst{2,3}
	\and
	M.~Zechmeister\inst{1}
}

\institute{Institut f\"ur Astrophysik, Georg-August-Universit\"at, Friedrich-Hund-Platz 1, 37077 G\"ottingen, Germany\\
	\email{david.cont@uni-goettingen.de}
	\and
	Instituto de Astrof{\'i}sica de Canarias (IAC), Calle V{\'i}a Lactea s/n, 38200 La Laguna, Tenerife, Spain
	\and
	Departamento de Astrof{\'i}sica, Universidad de La Laguna, 38026  La Laguna, Tenerife, Spain
	\and
	Leiden Observatory, Universiteit Leiden, Postbus 9513, 2300 RA, Leiden, The Netherlands
	\and
	Max-Planck-Institut f{\"u}r Astronomie, K{\"o}nigstuhl 17, 69117 Heidelberg, Germany
	\and
	Hamburger Sternwarte, Universit{\"a}t Hamburg, Gojenbergsweg 112, 21029 Hamburg, Germany
	\and
	Instituto de Astrof{\'i}sica de Andaluc{\'i}a (IAA-CSIC), Glorieta de la Astronom{\'i}a s/n, 18008 Granada, Spain
	\and
	European Southern Observatory (ESO), Alonso de C{\'o}rdova 3107, Vitacura, Casilla 19001, Santiago de Chile, Chile
	\and
	Institut de Ci\`encies de l'Espai (CSIC-IEEC), Campus UAB, c/ de Can Magrans s/n, 08193 Bellaterra, Barcelona, Spain
	\and
	Institut d'Estudis Espacials de Catalunya (IEEC), 08034 Barcelona, Spain
	\and
	Landessternwarte, Zentrum f\"ur Astronomie der Universit\"at Heidelberg, K\"onigstuhl 12, 69117 Heidelberg, Germany
	\and
	Centro de Astrobiolog{\'i}a (CSIC-INTA), ESAC, Camino bajo del castillo s/n, 28692 Villanueva de la Ca{\~n}ada, Madrid, Spain
	\and	
	Departamento de F\'{i}sica de la Tierra y Astrof\'{i}sica 
	and IPARCOS-UCM (Instituto de F\'{i}sica de Part\'{i}culas y del Cosmos de la UCM), 
	Facultad de Ciencias F\'{i}sicas, Universidad Complutense de Madrid, E-28040, Madrid, Spain
	\and
	Th{\"u}ringer Landessternwarte Tautenburg, Sternwarte 5, 07778 Tautenburg, Germany
	\\      
}
\date{Received 5 March 2021 / Accepted 14 May 2021}


\abstract
{Theoretical studies predict the presence of thermal inversions in the atmosphere of highly irradiated gas giant planets. Recent observations have identified these inversion layers. However, the role of different chemical species in their formation remains unclear.}
{We search for the signature of the thermal inversion agents TiO and Fe in the dayside emission spectrum of the ultra-hot Jupiter WASP-33b.}
{The spectra were obtained with CARMENES and HARPS-N, covering different wavelength ranges. Telluric and stellar absorption lines were removed with \texttt{SYSREM}. We cross-correlated the residual spectra with model spectra to retrieve the signals from the planetary atmosphere.}
{We find evidence for TiO at a significance of 4.9$\mathrm{\sigma}$ with CARMENES. The strength of the TiO signal drops close to the secondary eclipse. No TiO signal is found with HARPS-N. An injection-recovery test suggests that the TiO signal is below the detection level at the wavelengths covered by HARPS-N. The emission signature of Fe is detected with both instruments at significance levels of 5.7$\mathrm{\sigma}$ and 4.5$\mathrm{\sigma}$, respectively. By combining all observations, we obtain a significance level of 7.3$\mathrm{\sigma}$ for Fe. We find the TiO signal at $K_\mathrm{p}$\,=\,$248.0_{-2.5}^{+2.0}$\,km\,s$^{-1}$, which is in disagreement with the Fe detection at $K_\mathrm{p}$\,=\,$225.0_{-3.5}^{+4.0}$\,km\,s$^{-1}$. The $K_\mathrm{p}$ value for Fe is in agreement with prior investigations. The model spectra require different temperature profiles for TiO and Fe to match the observations. We observe a broader line profile for Fe than for TiO.}
{Our results confirm the existence of a temperature inversion layer in the planetary atmosphere. The observed $K_\mathrm{p}$ offset and different strengths of broadening in the line profiles suggest the existence of a TiO-depleted hot spot in the planetary atmosphere.}

\maketitle

%

\section{Introduction}

Hot Jupiters are gas giant planets with close-in orbits, which exhibit strong spectral features due to their enhanced temperatures and sizes. So far, targets of atmospheric observations have been mostly among this class of exoplanets. \cite{Hubeny2003} and \cite{Fortney2008} predicted the existence of temperature inversion layers in highly irradiated gas planets due to the strong absorption of visible and UV radiation caused by TiO and VO in the upper atmosphere. Initial evidence for the presence of an atmospheric inversion was found in the spectrum of the hot Jupiter HD 209458b by \cite{Knutson2008}. However, the claim of an inverted atmospheric temperature profile could not be confirmed \citep{Hansen2014, Diamond-Lowe2014, Schwarz2015, Evans2015}, and \cite{Hoeijmakers2015} did not find any signatures of TiO in high-resolution spectra of the planet. Moreover, \cite{Hoeijmakers2015} investigated the TiO line database and found several wavelength ranges with poor line list precision.

Producing high-resolution spectral line lists for transition metal diatomic molecules is computationally challenging, but the completeness and accuracy of line lists are critical for the detection of these chemical species \citep{McKemmish2019, Merritt2020}. Nevertheless, hints for the existence of TiO and VO were found in the atmospheres of WASP-33b and WASP-121b in secondary eclipse measurements by \cite{Haynes2015} and \cite{Evans2017}, respectively. High-resolution Doppler spectroscopy led to the detection of TiO in the emission spectrum of WASP-33b \citep{Nugroho2017}. However, \cite{Herman2020} reported a non-detection of TiO at high spectral resolution. More recently, \cite{Serindag2021} reassessed the presence of TiO in the spectra from \cite{Nugroho2017} by using an improved TiO line list, but they could not find a detection at the same orbital parameters as the previous work. Also the existence of TiO and VO in WASP-121b could not be confirmed by high-resolution spectroscopy observations \citep{Merritt2020}. TiO was also detected in the low-resolution transmission spectrum of WASP-19b by \cite{Sedaghati2017}, but the detection was not confirmed by \cite{Espinoza2019}. 

To explain these non-detections of metal oxides in the atmospheres of highly irradiated planets, a number of mechanisms have been proposed. A theoretical study by \cite{Spiegel2009} predicted the depletion of TiO and VO in the upper atmosphere of hot Jupiters, as gravitational settling moves the species into deeper layers of the atmosphere. Close-in giant planets are assumed to be tidally locked, with a permanent day- and nightside. Hence, the existence of a day-night cold-trap effect has been suggested. In this scenario, TiO and VO are moved by winds on a global scale to the nightside of the planet \citep{Parmentier2013}, where they condense due to cooler temperatures and they are thus efficiently removed. According to the studies of \cite{Lothringer2018} and \cite{Lothringer2019}, temperature inversions are also sensitive to the spectral type of the host star. Their simulations predict the occurrence of thermal dissociation and ionization in planetary atmospheres around hot stars, which decrease molecular abundances in favor of atomic species and ions.

Although TiO and VO have not been widely detected in hot Jupiters, thermal inversions have been found in several planets, such as WASP-33b \citep{Haynes2015, Nugroho2017, Nugroho2020b}, WASP-121b \citep{Evans2017}, WASP-19b \citep{Sedaghati2017}, WASP-18b \citep{Sheppard2017, Arcangeli2018}, WASP-103b \citep{Kreidberg2018}, and HAT-P-7b \citep{Mansfield2018}. These planets are all ultra-hot Jupiters (UHJs), that is gas giant planets with dayside temperatures greater than 2200\,K \citep{Parmentier2018}. Theoretical simulations \citep[e.g.,][]{Lothringer2018} suggest that the absorption of atoms and ions can produce thermal inversion layers in UHJs. Extreme thermal conditions lead to the dissociation of molecules into their constituent elements, allowing us to characterize the elemental composition of UHJ atmospheres.

Recently, atomic hydrogen was found in the transmission spectra of UHJs \citep[e.g.,][]{Yan2018, Casasayas-Barris2018, Jensen2018, Cauley2021, Yan2021}. Moreover, metals such as Fe, Mg, Na, Ca, Ti, or V and their ions were detected via transmission spectroscopy in the atmospheres of KELT-9b, KELT-20b, WASP-121b, WASP-12b, and WASP-33b \citep[e.g.,][]{Fossati2010, Hoeijmakers2018, Casasayas-Barris2019, Hoeijmakers2019, Cauley2019, Sing2019, Yan2019, Stangret2020, Nugroho2020a, Gibson2020, BenYami2020, Hoeijmakers2020}. For some planets, the detected spectral lines allowed for properties of their atmospheres to be analyzed in more detail, including atmospheric mass loss rate \citep[e.g.,][]{Yan2018, Wyttenbach2020} and nightside condensation \citep{Ehrenreich2020, Kesseli2021}. In addition to the transmission spectra, emission features of neutral iron consistent with the presence of an inversion layer were observed in the dayside spectra of the UHJs KELT-9b, WASP-189b, and WASP-33b \citep{Pino2020, Yan2020, Nugroho2020b}.
These detections of atoms and ions, together with the absence of TiO and VO in the spectra of several UHJs suggest that metals are likely to play a key role in the formation of thermal inversions.

WASP-33b (planet and host star parameters are listed in Table \ref{tab-parameters}) moves on a retrograde orbit around a $\delta$ Scuti A-type star with a period of 1.22\,days. The host star has a visual magnitude of $V$\,$\sim$\,~8\,mag, which makes WASP-33b a favorable target for observations. With an equilibrium temperature ($T_\mathrm{eq}$) of 2700\,K and a dayside temperature of $T_\mathrm{day}$\,$\sim$\,3000\,K, WASP-33b is the second hottest exoplanet known so far. This makes the planet an ideal candidate for investigating the role of chemical species in thermal inversions, their effect on the energy budget, and global circulation of strongly irradiated atmospheres. WASP-33b shows evidence for the presence of an inversion layer \citep{Haynes2015}. In addition to the detection of TiO by \cite{Nugroho2017}, \cite{Yan2019} found \ion{Ca}{ii} and \cite{Nugroho2020b} detected the presence of Fe as well as the existence of a thermal inversion via high-resolution Doppler spectroscopy. Hints for other high temperature absorption species were also found by \cite{Essen2019} and \cite{Kesseli2020}, who tentatively detected the spectral signature of AlO and FeH, respectively.

\begin{table}
	\caption{Parameters of the \object{WASP-33} system used in this work.}             
	\label{tab-parameters}                           
	\centering                                       
	\renewcommand{\arraystretch}{1.2} 
	\begin{threeparttable}
		\begin{tabular}{l l l}                       
			\noalign{\smallskip}
			\hline\hline                             
			\noalign{\smallskip}
			Parameter & Symbol [Unit] & Value \\     
			\noalign{\smallskip}
			\hline                                   
			\noalign{\smallskip}
			\textit{Planet} & &  \\ 
			\noalign{\smallskip}
			Radius  \tablefootmark{a}               & $R_\mathrm{p}$ [$R_\mathrm{J}$] & $1.679_{-0.030}^{+0.019}$ \\
			Orbital period \tablefootmark{b}      & $P_\mathrm{orb}$ [d] & 1.219870897 \\
			Transit epoch (BJD) \tablefootmark{b}  & $T_\mathrm{0}$ [d] & 2454163.22449\\
			Systemic velocity \tablefootmark{c}    & $\varv_\mathrm{sys}$ [km\,s$^{-1}$] & $-3.0\pm0.4$\\
			RV semi-amplitude \tablefootmark{a} & $K_\mathrm{p}$ [km\,s$^{-1}$] & $231\pm3$\\
			Duration ingress \tablefootmark{d}     & $T_\mathrm{ingress}$ [d] & $0.0124\pm0.0002$\\
			Duration transit \tablefootmark{d}     & $T_\mathrm{transit}$ [d] & $0.1143\pm0.0002$\\
			Surface gravity \tablefootmark{d}     & log $g$ [cgs] & 3.46\\
			\noalign{\smallskip} \hline \noalign{\smallskip}
			\textit{Star} & &  \\  
			\noalign{\smallskip}
			Radius \tablefootmark{a} & $R_*$ [$R_\mathrm{\sun}$] & $1.509_{-0.027}^{+0.016}$\\ 
			Effective temperature \tablefootmark{e} & $T_\mathrm{eff}$ [K] & $7430\pm100$\\			
			Rotational velocity \tablefootmark{f} & $\varv_\mathrm{rot}\sin i_*$ &  $86.63_{-0.32}^{+0.37}$\\ 
			& [km\,s$^{-1}$] & \\
			\noalign{\smallskip}
			\hline                                   
		\end{tabular}
		\tablefoot{
			\tablefoottext{a}{\cite{Lehmann2015} with parameters from \cite{Kovacs2013}}\\
			\tablefoottext{b}{\cite{Maciejewski2018}}\\
			\tablefoottext{c}{\cite{Nugroho2017}}\\
			\tablefoottext{d}{\cite{Kovacs2013}}\\
			\tablefoottext{e}{\cite{Cameron2010}}\\
			\tablefoottext{f}{\cite{Johnson2015}}			
		}
	\end{threeparttable}      
\end{table}

In this work, we report the detection of Fe and evidence for the presence of TiO on the dayside of WASP-33b. We use high-resolution Doppler spectroscopy with CARMENES (Calar Alto high-Resolution search for M dwarfs with Exoearths with Near-infrared and optical \'Echelle Spectrographs) and HARPS-N (High Accuracy Radial velocity Planet Searcher for the Northern hemisphere). The signature of both species is observed in emission, indicating the presence of an inverted temperature profile in the planetary atmosphere. We structure the paper as follows. In Section \ref{Observations} and Section \ref{Data reduction}, we describe our observations and the data reduction procedures. Section \ref{Methods} details the methodology used to find the signals of TiO and Fe. In Section \ref{Results and discussion}, we present the results with discussions. Conclusions are drawn in Section~\ref{Conclusions}.

%

\section{Observations}
\label{Observations}

%
\begin{table*}
	\caption{Observation log. The observations from CARMENES and HARPS-N are presented the first time in this work (new data). The observations with ESPaDOnS are archival data from \cite{Herman2020}.}             
	\label{obs_log}      
	\centering                          
	\begin{threeparttable}
		\begin{tabular}{l l l l l l l }        
			\hline\hline                 
			\noalign{\smallskip}
			Instrument & Date & Observing time & Airmass change & Phase coverage & Exposure time & $N_\mathrm{spectra}$  \\     
			\noalign{\smallskip}
			\hline                       
			\noalign{\smallskip}
			\textit{New data} & &  \\  
			\noalign{\smallskip}
			CARMENES & 2017-11-15 & 17:59--04:47\,UT	 & 1.87--1.00--2.53 &  0.29--0.65  & 300\,s & 105\\  
			HARPS-N  & 2020-10-15 & 21:06--04:55\,UT	 & 1.99--1.01--1.27 &  0.43--0.70  & 200\,s & 125\\  
			HARPS-N & 2020-11-07 & 19:39--05:18\,UT	 & 1.96--1.01--2.03 &  0.24--0.57  & 200\,s & 155\\
			\noalign{\smallskip} \hline \noalign{\smallskip}
			\textit{Archival data} & &  \\  
			\noalign{\smallskip}
			ESPaDOnS & 2013-09-15 & 09:09--13:00\,UT	 & 1.75--1.05 &  0.30--0.44  & 90\,s & 110\\  
			ESPaDOnS & 2013-09-26 & 10:35--12:29\,UT	 & 1.16--1.05 &  0.37--0.44  & 90\,s & 55\\  
			ESPaDOnS & 2014-09-04 & 10:49--14:42\,UT	 & 1.39--1.05--1.07 &  0.56--0.69  & 90\,s & 110\\  
			ESPaDOnS & 2014-09-15 & 10:00--13:53\,UT	 & 1.42--1.05--1.06 &  0.55--0.68  & 90\,s & 110\\  
			ESPaDOnS & 2014-11-05 & 08:51--10:49\,UT	 & 1.08--1.05--1.08 &  0.31--0.38  & 90\,s & 55\\  			
			\noalign{\smallskip}
			\hline                                   
		\end{tabular}
	\end{threeparttable}      
\end{table*}

\begin{figure}
	\centering
	\includegraphics[width=0.5\textwidth]{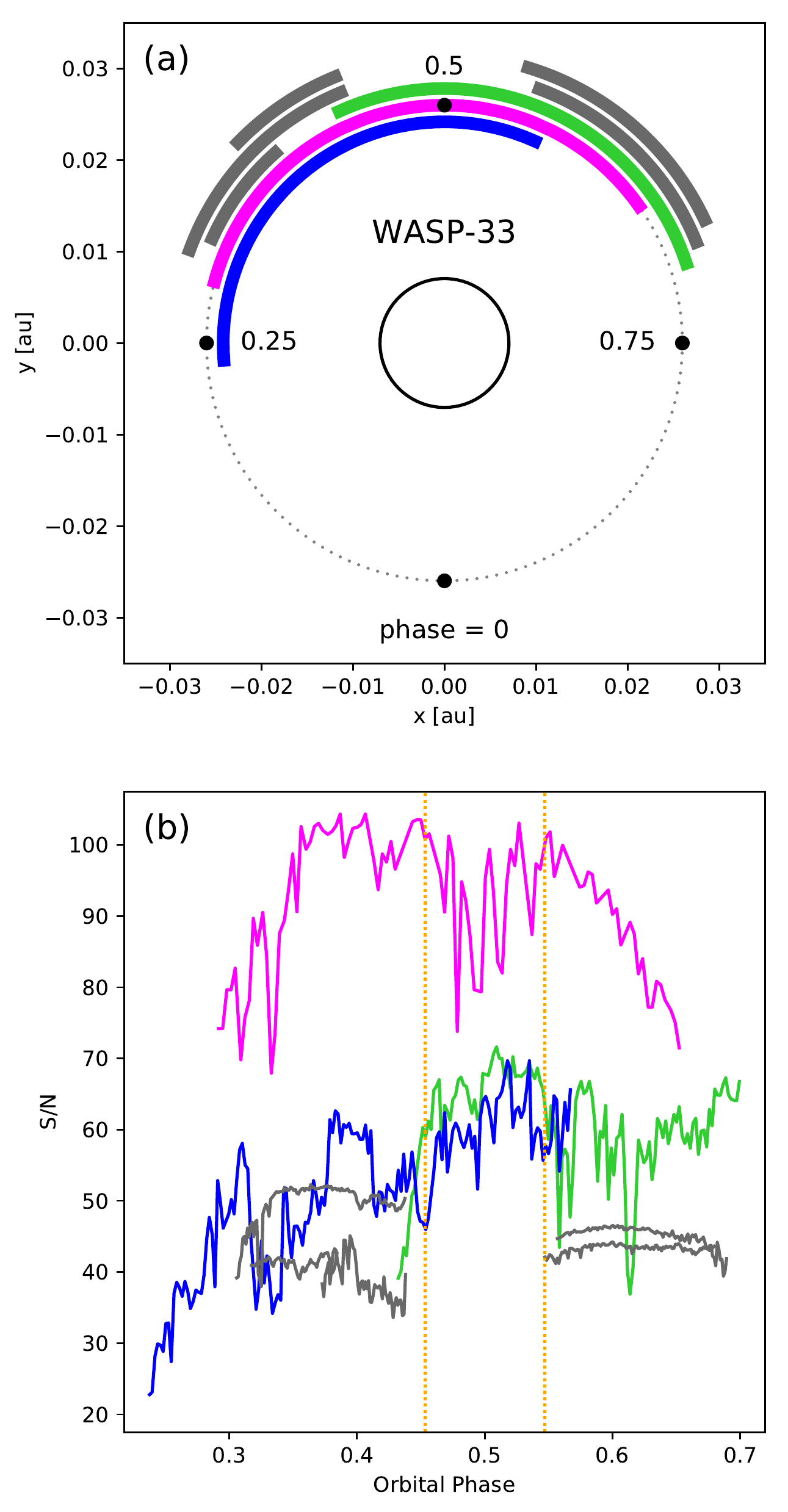}
	\caption{Orbital phase coverage and S/N. Pink corresponds to CARMENES, blue/green to HARPS-N and gray to the ESPaDOnS observations \citep{Herman2020}. Panel \textit{(a)} is the WASP-33 system with the phase coverage of the observations. Panel \textit{(b)} is the S/N of each spectrum as a function of orbital phase. The begin and the end of the secondary eclipse are indicated by the yellow dashed lines.}
	\label{orbital-phase-coverage-SNR}
\end{figure}

We observed the thermal emission spectrum of WASP-33b on 15 November 2017 with the CARMENES spectrograph \citep{Quirrenbach2016, Quirrenbach2018, Quirrenbach2020} at the 3.5\,m Calar Alto telescope. CARMENES consists of two fiber fed high-resolution spectrographs covering the wavelength ranges from 520 to 960\,nm (VIS) and from 960 to 1710\,nm (NIR), which corresponds to 61 and 28 spectral orders, respectively. The resolution is $R$\,$\sim$\,94,600 in the VIS channel and $R$\,$\sim$\,80,400 in the NIR channel. In this work, only the data collected with the VIS channel are used. The CARMENES observation covered pre-eclipse, eclipse and post-eclipse of the planet, which corresponds to an orbital phase coverage of 0.29 to 0.65 (cf. Fig.~\ref{orbital-phase-coverage-SNR}a). In total, we gathered 105 spectra, each with an exposure time equal to 300\,s. The airmass varied between 1.00--2.53. Except for one thick cirrus, the night was photometric with a seeing of about 1.5\,$\arcsec$. We discarded seven spectra for which the target was not centered onto the fiber due to bad guiding of the telescope. Moreover, we removed three spectra during the passing of the cirrus cloud, ending up with a total number of 95 spectra for further analysis.

Another two observations of the thermal emission spectrum of WASP-33b were obtained on 15 October 2020 and 7 November 2020 with the HARPS-N spectrograph \citep{Mayor2003, Cosentino2012} at the Telescopio Nazionale \textit{Galileo}. HARPS-N is a fiber fed high-resolution spectrograph that covers the wavelength range from 383 to 690\,nm, corresponding to 69 spectral orders. The spectral resolution is $R$\,$\sim$\,115,000. Our observations covered the orbital phase range 0.43 to 0.70 in the first night and 0.24 to 0.57 in the second night (cf. Fig.~\ref{orbital-phase-coverage-SNR}a). We obtained 125 and 155 spectra, respectively. The exposure time of each spectrum was 200\,s for both observations. The airmass varied between 1.01--1.99 and 1.01--2.03, respectively.

For all the observations, we observed the target with fiber A and used fiber B to record the sky background for considering the sky emission lines in the subsequent data analysis. Further details on the observations are reported in the observation log in Table \ref{obs_log}.

In addition to our data from CARMENES and HARPS-N, we also re-analyzed five archival observations from ESPaDOnS (Echelle SpectroPolarimetric Device for the Observation of Stars) at the Canada-France-Hawaii telescope \citep{Herman2020}. Details of this analysis are provided in Sections \ref{Comparison with previous work} and \ref{Combining the Fe signal}.

%

\section{Data reduction}
\label{Data reduction}

\subsection{Preprocessing the spectra}
\label{Pre-processing the spectra}

\begin{figure*}
	\centering
	\includegraphics[width=\textwidth]{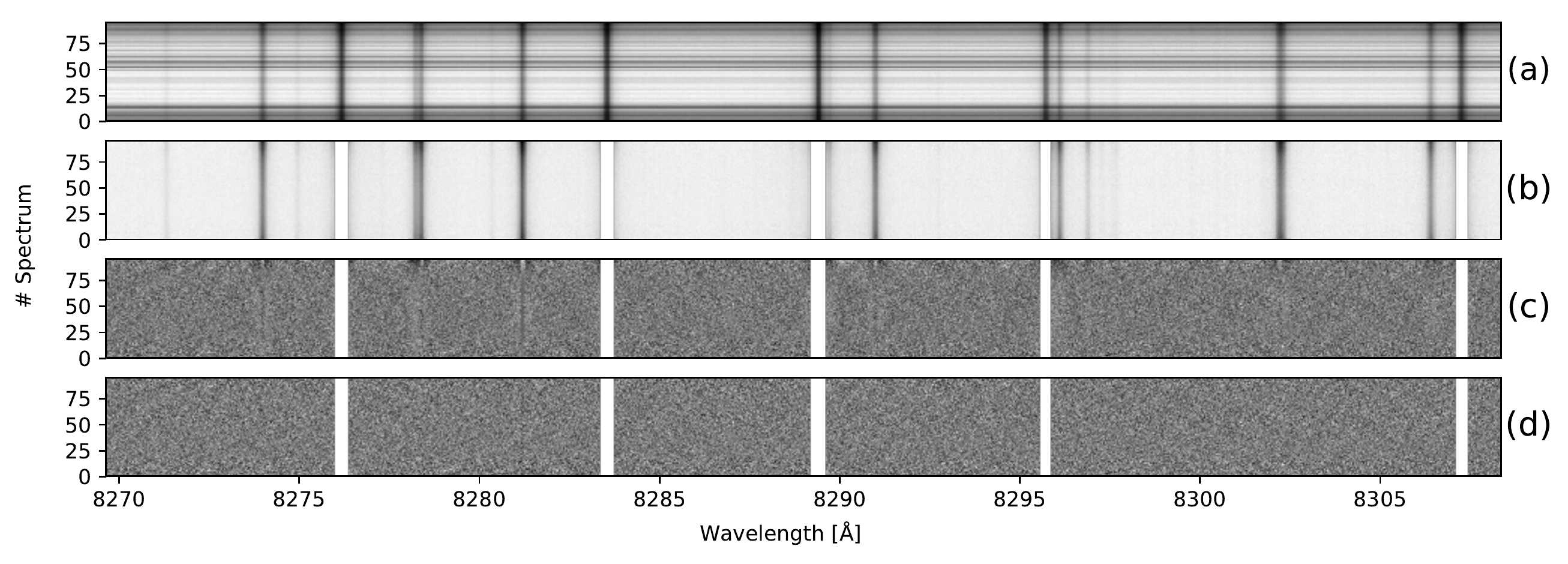}
	\caption{Preprocessing steps for a selected wavelength range of CARMENES. Panel \textit{(a)} is the unprocessed spectral matrix; panel \textit{(b)} is the matrix after normalization, masking and outlier correction; panels \textit{(c)} and \textit{(d)} are the spectral residuals after one and six \texttt{SYSREM} iterations, respectively. After one iteration, several telluric lines are still visible. At higher iteration numbers the telluric lines are almost entirely removed.}
	\label{spectra-raw-normalized-masked-sysrem}
\end{figure*}

The raw frames were processed by the data reduction pipelines CARACAL v2.20 for CARMENES \citep{Zechmeister2014, Caballero2016} and the Data Reduction Software for HARPS-N. With CARMENES, we obtained a one-dimensional spectrum for each frame and spectral order (our numbering is 1--61, corresponding to the CARMENES echelle orders 118--58). We split the order-merged, one-dimensional spectra from HARPS-N into 69 order-like segments to conduct the same analysis steps for both instruments (hereafter spectral orders; for wavelength range of each segment see Fig.~\ref{CCF-with-Barnards-star_HARPS}). The flux signal-to-noise ratio (S/N) of each spectral order and one-dimensional spectrum was calculated by the instrument pipelines. We report the mean S/N value of each spectrum in Fig.~\ref{orbital-phase-coverage-SNR}b.

We used Python to apply the following procedures to the spectra. After sorting the spectra chronologically, we obtained a two dimensional matrix for each observation and spectral order (see an example of spectral matrix in Fig.~\ref{spectra-raw-normalized-masked-sysrem}a). We corrected pixels that the pipelines flagged as bad quality by linear interpolation to the nearest neighbors. Pixels that were flagged more than three times during the time series were masked. To correct 5$\sigma$ outliers due to cosmic rays, we fit a third order polynomial to the time evolution of each pixel and replaced the affected pixels with the polynomial function values. Furthermore, we needed to remove the contribution of the grating blaze function and the different exposure levels of the spectra due to the varying atmospheric conditions (e.g., changing airmass) during the observations. Hence, we individually fit a second order polynomial to the pseudo-continuum of each spectrum and normalized it with the resulting fit function. Wavelength ranges with broad stellar absorption bands or strong emission lines in fiber~B were excluded during the second order polynomial fit. We masked wavelength ranges where the flux was below 20\% of the continuum level. Due to almost no flux, the five spectral orders at the red end of the CARMENES wavelength range were entirely masked and excluded from further analysis. As a result, we obtained a normalized, masked and outlier corrected spectral matrix for each observation and spectral order (see example in Fig.~\ref{spectra-raw-normalized-masked-sysrem}b). 

\begin{figure*}
	\centering
	\includegraphics[width=\textwidth]{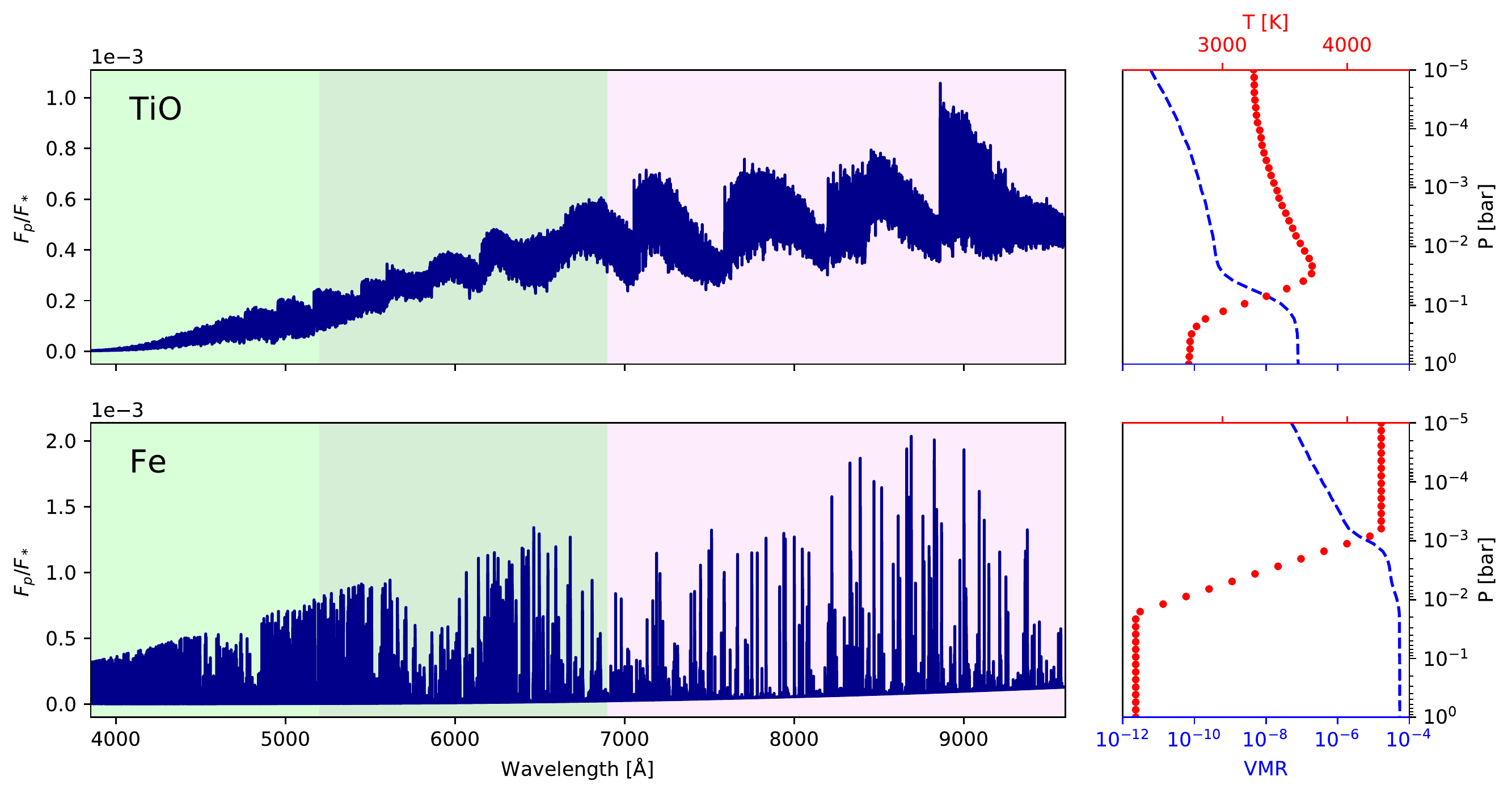}
	\caption{Modeled planetary emission spectra ({\it left panels}) and the corresponding $T$-$p$ profiles with volume mixing ratios (VMRs, {\it right panels}). We used different $T$-$p$ profiles for the two species and computed the VMRs assuming equilibrium chemistry with solar elemental abundances. The upper and the lower panels refer to TiO and Fe, respectively. The CARMENES wavelength range is shaded in pink; the HARPS-N wavelength range is shaded in green; the ranges overlap. The spectral lines are stronger in the CARMENES range when compared to HARPS-N. A different scaling is applied to the planet-to-star flux contrast ratio on the y-axis of the spectra. The spectral lines of Fe are stronger when compared with the TiO model spectrum.}
	\label{T-P-profiles-and-template-spectra}
\end{figure*}

\subsection{Removal of telluric and stellar lines}
\label{Removal of telluric and stellar lines}
The Earth's telluric and stellar lines were removed from the spectra by using \texttt{SYSREM}, a detrending algorithm originally designed to remove systematic effects from sets of transit light curves \citep{Tamuz2005}. The algorithm iteratively performs linear fits of the stellar and telluric line evolution in time and then subtracts the linear contribution from the signal. We treated each wavelength bin as an individual light curve to remove systematics from the spectral time series. The spectra were detrended by passing each two dimensional spectral matrix as input to the algorithm. We assigned airmass as the starting fit parameter in order to improve the performance of \texttt{SYSREM}. We ran the algorithm for different iteration numbers, that is between one and ten consecutive iterations (see example of different iterations in Fig.~\ref{spectra-raw-normalized-masked-sysrem}c and Fig.~\ref{spectra-raw-normalized-masked-sysrem}d). To remove large-scale features, we filtered the resulting spectral residual matrices with a Gaussian high-pass filter (25\,pixels for CARMENES; 75\,pixels for HARPS-N) and divided each matrix column by its variance \citep{Yan2019}.

When assuming a circular orbit with a semi-amplitude velocity equal to 231\,km\,s$^{-1}$ \citep{Lehmann2015,Kovacs2013}, the planet is expected to move at radial velocities between --220\,km\,s$^{-1}$ and +231\,km\,s$^{-1}$ during our observations (phase coverage 0.24--0.70). On the other hand, the telluric and stellar lines are approximately stationary. Therefore, at small iteration numbers, the \texttt{SYSREM} algorithm removes mostly the telluric and stellar lines while only slightly affecting the planetary signal. However, once telluric and stellar lines are fit and removed to a certain degree, the algorithm begins to detrend also the planetary lines \citep[e.g.,][]{Birkby2017, Nugroho2017, Alonso-Floriano2019, Sanchez-Lopez2019}. Hence, we expect the planetary signal to appear strongest after an optimal number of \texttt{SYSREM} iterations. This number should vary from order to order due to a different strength of telluric and stellar line contamination. However, we decided to use a conservative approach and assumed a common optimal iteration number for all spectral orders. The amplitude of the signal from the planetary atmosphere varies between different wavelength ranges and chemical species. Therefore, we assessed the optimal iteration number by maximizing the detection strength for each instrument and chemical species separately (see Sections \ref{Detection of Fe} and \ref{Evidence for TiO}).


\section{Methods}
\label{Methods}

The planetary signal is buried in the noise of the residual matrices (see example in Fig.~\ref{spectra-raw-normalized-masked-sysrem}d). To extract the atmospheric emission signature, we employed the cross-correlation method, which has been successfully applied in a number of previous studies \citep[e.g.,][]{Snellen2010, Brogi2012, Rodler2012, Birkby2013, Snellen2014, Alonso-Floriano2019, Sanchez-Lopez2019}. This technique allows us to combine the numerous weak planetary lines into a detectable signal by computing the cross-correlation function (CCF) between the residual spectra and a planetary model spectrum. To this end, we computed model spectra for the chemical species we intended to search for (i.e., TiO and Fe).

\subsection{Spectral models}
\label{Spectral models}

Molecular species are affected by thermal dissociation in the dayside atmosphere of UHJs \citep[e.g.,][]{Kreidberg2018, Parmentier2018, Arcangeli2019}. As TiO may be depleted at the locations where the temperature is highest (near the substellar point), we hypothesize that the spectral signatures of TiO and Fe may emerge from atmospheric regions with different thermal conditions. For this reason, we decided to model two different atmospheres, each with an individual thermal structure. Both atmospheres consist of 40 layers in a pressure range from $10^{-5}$ to 1\,bar and are equispaced on a logarithmic scale. We assumed a moderate temperature for the TiO atmosphere (to avoid TiO to be significantly dissociated) and a higher temperature for Fe. In Section \ref{Discussion}, we further discuss the usage of two different temperature profiles. For TiO, we took the inverted $T$-$p$ profile found by \cite{Haynes2015}. To model the Fe spectrum, we used the $T$-$p$ profile of WASP-189b, which was retrieved by \cite{Yan2020} using the Fe emission lines. This planet has properties similar to WASP-33b (e.g., mass, radius, equilibrium temperature, and spectral type of the host star). Therefore, using this $T$-$p$ profile is an appropriate approximation for studying the presence of Fe in the atmosphere of WASP-33b. We used \texttt{easyCHEM} \citep{Molliere2017} to compute the volume mixing ratio (VMR) and the mean molecular weight for each layer under the assumption of equilibrium chemistry and solar elemental abundances. We calculated emission spectra for TiO and Fe using the radiative transfer code \texttt{petitRADTRANS} \citep{Molliere2019}. We used a blackbody spectrum with a temperature of 7430\,K for the host star to compute the planet-to-star flux ratio. The flux normalized model spectra as well as the corresponding $T$-$p$ profiles and VMRs are shown in Fig.~\ref{T-P-profiles-and-template-spectra}. We convolved the normalized emission model spectra with the instrument profiles and applied the same high-pass filter as described for the residual spectra in Section \ref{Removal of telluric and stellar lines}. This makes our analysis insensitive to the emission continuum level and consequently, allows us to only account for the relative strength of the spectral lines.

\subsection{Cross-correlation}
\label{Cross-correlation}
Our implementation of the cross-correlation is based on the Python routine \texttt{pyasl.crosscorrRV} from the \texttt{PyAstronomy} package \citep{Czesla2019}. We computed the CCFs over a range of Doppler shifts from --364\,km\,s$^{-1}$ to +364\,km\,s$^{-1}$ and applied velocity steps of 1.3\,km\,s$^{-1}$ for CARMENES and 0.8\,km\,s$^{-1}$ for HARPS-N , which corresponds to the mean pixel spacing of the instruments. A CCF with the planet model spectrum was calculated for each residual spectrum. This resulted in a 95$\times$561 cross-correlation matrix ($\overline{\mathrm{CCF}}$) for CARMENES and to a 125$\times$911 and a 155$\times$911 $\overline{\mathrm{CCF}}$ for HARPS-N. We subtracted the median value from each CCF to avoid any interference with leftover broadband features in the spectra. The $\overline{\mathrm{CCF}}$s were calculated independently for all spectral orders, chemical species and observations.

\begin{figure}
	\centering
	\includegraphics[width=0.5\textwidth]{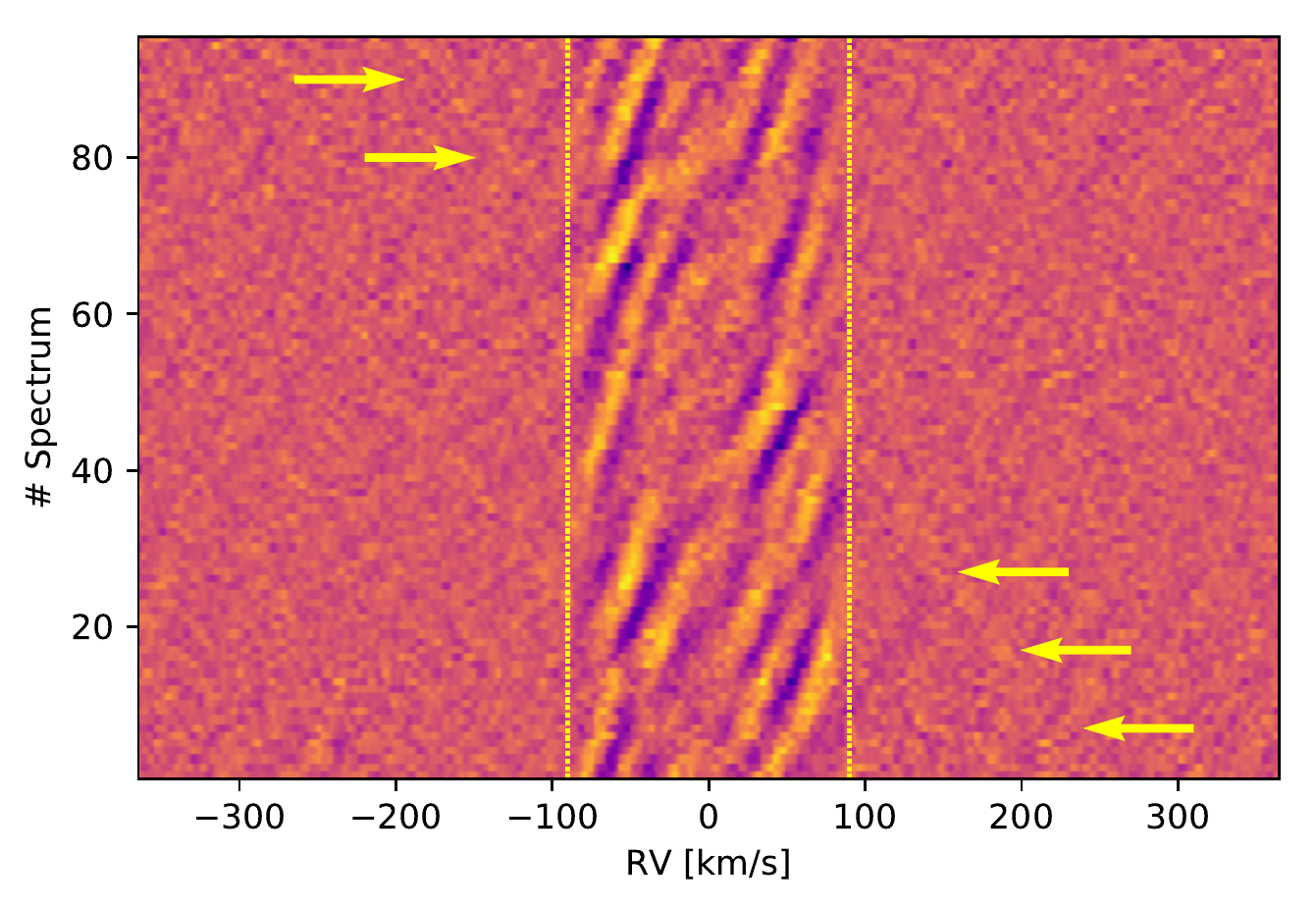}
	\caption{CCF map of Fe obtained with CARMENES in the stellar rest frame. The strong residuals between the yellow dashed lines are caused by residual Fe lines from the host star pulsation. We indicate the faint planetary trail with yellow arrows.}
	\label{CCF-map_CARMENES_Fe}
\end{figure}

We co-added the $\overline{\mathrm{CCF}}$s for each chemical species and observation separately. This led to the final cross-correlation maps. Only the $\overline{\mathrm{CCF}}$s of the spectral orders that are selected in Section \ref{Exclusion of bad spectral orders} were included in this step. This resulted in a final cross-correlation function map (CCF map) for each chemical species and observation (e.g., Fig.~\ref{CCF-map_CARMENES_Fe}). Finally, we merged the CCF maps of the two HARPS-N observations into one single 280$\times$911 CCF map, which led to one CCF map for each instrument and chemical species.

We found strong artifacts in the CCF map of Fe that originate from residual stellar lines (see Fig.~\ref{CCF-map_CARMENES_Fe}). These lines are not efficiently removed by \texttt{SYSREM} as their strength varies with time due to the pulsation of the host star. The radial velocity (RV) domain of the residual stellar lines is determined by the stellar rotation velocity and is confined to the range between $\pm \varv_\mathrm{rot}\sin i_*$ (i.e., between roughly --87\,km\,s$^{-1}$ and +87\,km\,s$^{-1}$ in the stellar rest frame). To avoid any correlation with the model spectrum in the CCF map of Fe, we masked all velocities between \mbox{--90\,km\,s$^{-1}$} and +90\,km\,s$^{-1}$ in the stellar rest frame (for comparison, the planetary RV during secondary eclipse is between --67\,km\,s$^{-1}$~and~+67\,km\,s$^{-1}$). Hence, the RV ranges from \mbox{--90\,km\,s$^{-1}$}~to~--67\,km\,s$^{-1}$ and from +67\,km\,s$^{-1}$~to~+90\,km\,s$^{-1}$ are lost as the planetary trail in the CCF map is masked. This represents 10\% (CARMENES) and 16\% (HARPS-N) of the observations outside eclipse. Consequently, the masking of the stellar line residuals will not strongly affect the detection in Section \ref{Detection of Fe} and the resulting conclusions.

\begin{figure*}
	\centering
	\includegraphics[width=\textwidth]{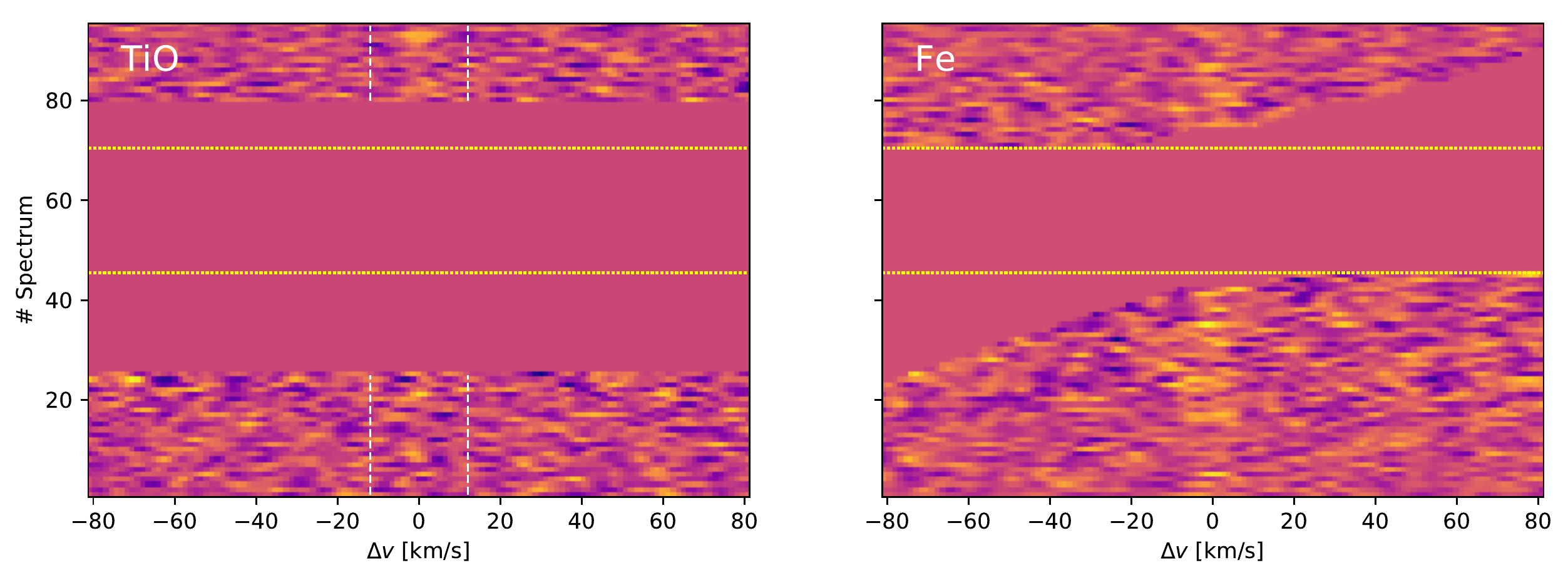}
	\caption{CCF maps from CARMENES aligned to the rest frame consistent with the maximum $K_\mathrm{p}$ values (i.e., 248.0\,km\,s$^{-1}$ for TiO and 228.0\,km\,s$^{-1}$ for Fe; see Fig.~\ref{SN_maps_CARMENES}). We indicate the TiO signature with the vertical dashed lines. The two horizontal dashed lines indicate the beginning and end of the secondary eclipse. We masked the phase interval where the TiO signal is below the noise level (see Section \ref{Evidence for TiO}). Also the radial velocity domain of residual stellar Fe lines was masked (see Section \ref{Cross-correlation}). The wider Fe trail when compared to TiO is likely caused by a different degree of rotational broadening, which hints at a global distribution of Fe and localized TiO in the atmosphere of WASP-33b.}
	\label{aligned-planetary-trails_CARMENES}
\end{figure*}
\begin{figure*}
	\centering
	\includegraphics[width=\textwidth]{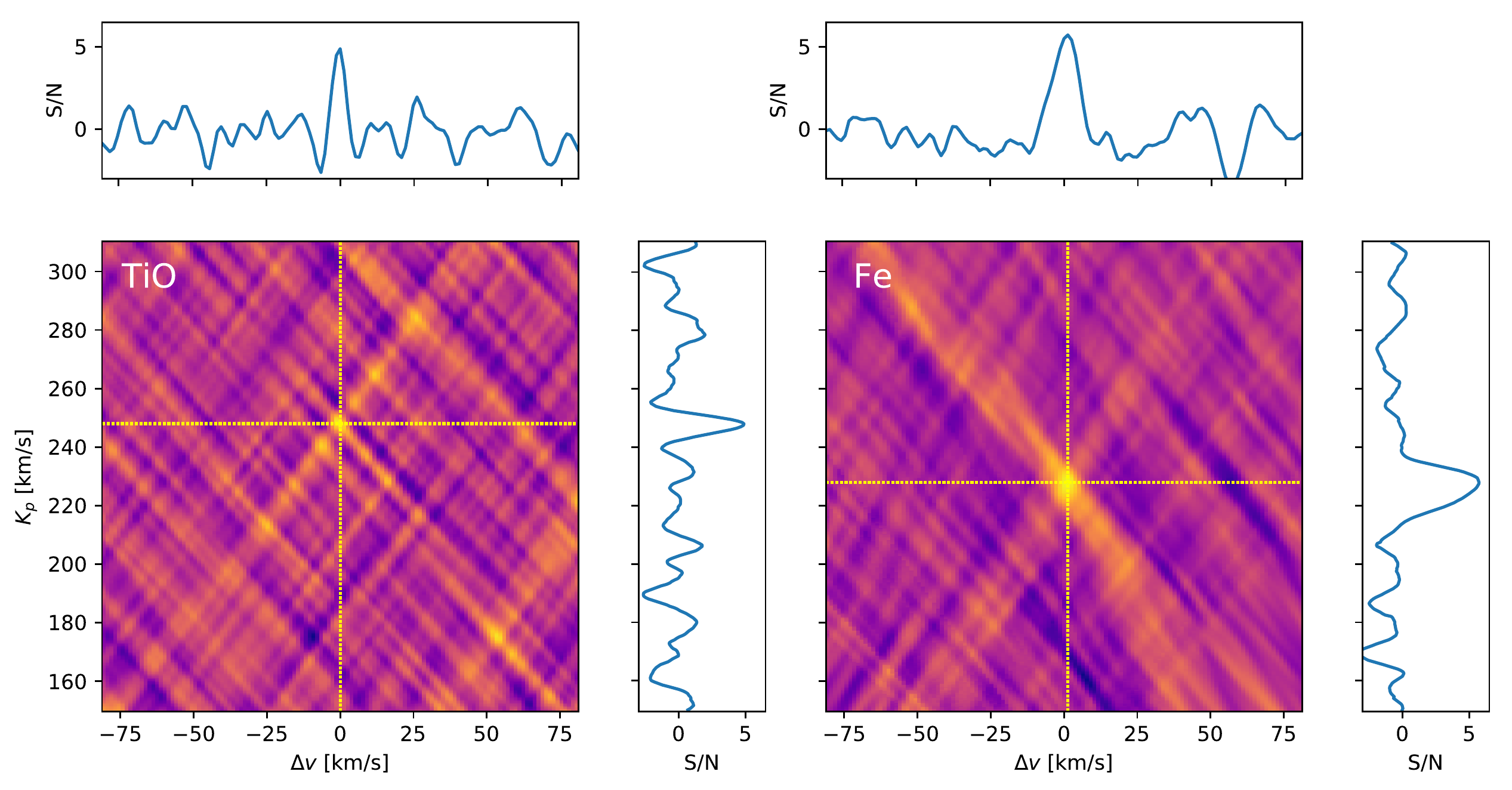}
	\caption{S/N maps after six and five consecutive \texttt{SYSREM} iterations with CARMENES for TiO and Fe, respectively. We get S/N significance levels of 4.9 for TiO and 5.7 for Fe. The peak coordinates in the S/N map are indicated by the yellow dashed lines. Cross-sections of the S/N peaks are reported in the horizontal and vertical panels. The horizontal panels also correspond to the collapsed CCF maps in Fig.~\ref{aligned-planetary-trails_CARMENES}. The Fe signal is consistent with the expected $K_\mathrm{p}$ value; the TiO peak is found with an offset of roughly +17\,km\,s$^{-1}$.}
	\label{SN_maps_CARMENES}
\end{figure*}

\begin{figure*}
	\centering
	\includegraphics[width=\textwidth]{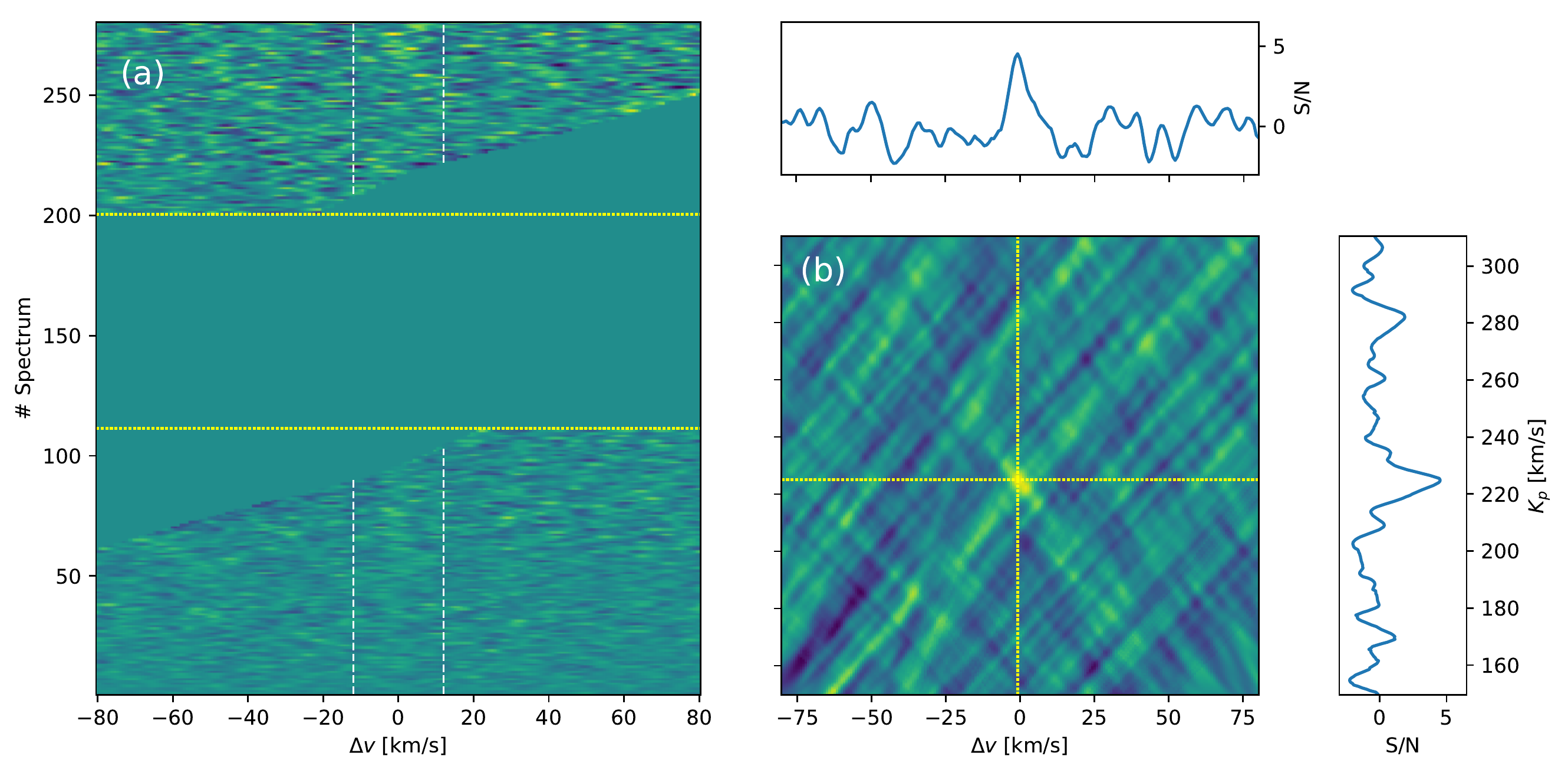}
	\caption{Detection of Fe with HARPS-N. \textit{(a)} CCF map in the planetary rest frame (aligned using $K_\mathrm{p}$\,=\,225.0\,km\,s$^{-1}$). The vertical trail indicated with white lines is the planetary Fe signal. The two horizontal dashed lines indicate the beginning and end of secondary eclipse. \textit{(b)} Signal-to-noise ratio map. The peak coordinates in the S/N map are indicated by the yellow dashed lines. Cross-sections of the S/N map are shown in the horizontal and vertical panels.}
	\label{aligned-planetary-trails_HARPS}
\end{figure*}

\subsection{Exclusion of bad spectral orders}
\label{Exclusion of bad spectral orders}
The precision of line lists is of critical importance when using the cross-correlation technique. However, the line lists of TiO suffer from inaccuracies, which reduce the detection sensitivity of the molecule \citep{Hoeijmakers2015, Nugroho2017}. The calculation of models leading to a detection of TiO in planetary atmospheres remains a challenging task \citep{Herman2020, Serindag2021}. We attempted to mitigate this issue by using the new line database \texttt{ToTo} \texttt{ExoMol} \citep{McKemmish2019} to generate our TiO model spectrum. However, this line list is also expected to show inaccuracies in certain wavelength ranges \citep{McKemmish2019}. To exclude spectral orders with a poor line list from our analysis, we conducted an order-wise validation of the TiO line list. A detailed description of the TiO line list validation is provided in Appendix~\ref{Validation of the TiO line list}. As a result, we only included the $\overline{\mathrm{CCF}}$s from spectral orders corresponding to wavelength ranges with a precise line list when generating the CCF map for TiO. In contrast, Fe line lists are considered to be precise \citep{Kurucz2011}. Hence, we refrained from analyzing the line list precision of Fe and included the $\overline{\mathrm{CCF}}$s of all orders to compute the CCF map of this species.

No prominent emission features are present in the TiO model spectrum blueward of about 6000\,\r{A} (see Fig.~\ref{T-P-profiles-and-template-spectra}). To assess whether the molecular signature in the corresponding spectral orders is strong enough to contribute to a detection of TiO, we conducted an order-wise injection-recovery test (Appendix \ref{Injection-recovery test}). Consequently, we included only the $\overline{\mathrm{CCF}}$s that allow us to recover an injected model spectrum into the CCF map. We recovered the injected model spectrum in several spectral orders of CARMENES successfully (roughly between 6000\,\r{A} and 9000\,\r{A}). In contrast, only in one spectral order of HARPS-N the injected signal could be retrieved. This result suggests that even if present, the signature of TiO will be below the required level for a significant detection in the HARPS-N observations. On the other hand, all the spectral orders contribute to the detection of Fe due to the stronger emission features in the model spectrum when compared to TiO (cf. Fig.~\ref{T-P-profiles-and-template-spectra}). Consequently, we included all the spectral orders when calculating the Fe CCF map.

In conclusion, only the spectral orders passing both the line list assessment and the injection-recovery test were included in our TiO analysis. All other spectral orders were excluded. In contrast, we included all spectral orders in the Fe analysis because of the availability of a precise line list and a strong spectral signature expected from the model spectrum in Fig.~\ref{T-P-profiles-and-template-spectra}.

\subsection{Searching for atmospheric features}
\label{Searching for atmospheric features}
We assumed that the planet moves on a circular orbit and expect its observed radial velocity to be described by the expression
\begin{equation}
\label{equ-orb-v}
\varv_\mathrm{p} = \varv_\mathrm{sys} + \varv_\mathrm{bary} + K_\mathrm{p} \sin\left(2\pi\phi\right) + \Delta \varv,
\end{equation}
where $\varv_\mathrm{sys}$ is the systemic velocity, $\varv_\mathrm{bary}$ is the observer's barycentric velocity, $K_\mathrm{p}$ is the orbital semi-amplitude velocity of the planet, $\phi$ is the orbital phase and $\Delta \varv$ is the residual radial velocity in the planetary rest frame. By using Eq.~(\ref{equ-orb-v}) and linear interpolation, we aligned the CCF map to the planetary rest frame (see Fig.~\ref{aligned-planetary-trails_CARMENES}). To account for the varying flux level from atmospheric conditions and instrumental effects (e.g., telescope guiding, alignment with the instrument fiber), each row of the CCF map was weighted with the squared flux S/N (see Fig.~\ref{orbital-phase-coverage-SNR}b) of the corresponding spectrum \citep{Brogi2019}. We collapsed the aligned CCF map along the time axis by computing the mean value of each matrix column. If the model spectrum reflects the planetary signal and a Doppler shift according to Eq.~(\ref{equ-orb-v}) is present, the collapsed CCF map will show a peak at $\Delta \varv$ close to 0\,km\,s$^{-1}$ (see top panels in Fig.~\ref{SN_maps_CARMENES}). Following the same procedure as previous studies \citep[e.g.,][]{Birkby2017, Sanchez-Lopez2019, Alonso-Floriano2019}, we aligned with different $K_\mathrm{p}$ values and combined the resulting 1D plots of the collapsed CCF map in a 2D matrix. We used $K_\mathrm{p}$ values between +150\,km\,s$^{-1}$ and +310\,km\,s$^{-1}$ in steps of 0.5\,km\,s$^{-1}$. We considered a range of $\Delta \varv$ from --80.6\,km\,s$^{-1}$ to +80.6\,km\,s$^{-1}$ in steps of 1.3\,km\,s$^{-1}$ for the CARMENES observation. The considered values for both HARPS-N observations ranged from --80.0\,km\,s$^{-1}$ to +80.0\,km\,s$^{-1}$ in steps of 0.8\,km\,s$^{-1}$. Under exclusion of the peak, we computed the standard deviation of the 2D matrix. The matrix was normalized with the standard deviation and a signal-to-noise map of the detection significance (S/N map) as a function of the orbital semi-amplitude $K_\mathrm{p}$ and the radial velocity deviation from the planetary rest frame $\Delta \varv$ was obtained. To assess the strength and the position of the detection peaks, we computed the S/N map from each instrument and chemical species independently.

%

\section{Results and discussion}
\label{Results and discussion}

\subsection{Detection of Fe}
\label{Detection of Fe}

We found a clear signature of Fe with both instruments. With CARMENES, we achieved a maximum peak value of S/N\,=\,5.7 at $K_\mathrm{p}$\,=\,$228.0_{-5.0}^{+3.5}$\,km\,s$^{-1}$ and $\Delta \varv$\,=\,$1.3_{-3.9}^{+3.9}$\,km\,s$^{-1}$ after five consecutive \texttt{SYSREM} iterations (see Fig.~\ref{SN_maps_CARMENES} right panel). For HARPS-N the maximum peak value of S/N\,=\,4.5 was found after eight iterations at $K_\mathrm{p}$\,=\,$225.0_{-5.0}^{+2.0}$\,km\,s$^{-1}$ and $\Delta \varv$\,=\,$-0.8_{-2.4}^{+4.0}$\,km\,s$^{-1}$ (see Fig.~\ref{aligned-planetary-trails_HARPS}b). A summary of all results is provided in Table \ref{tab-results}. The detected Fe signal is strong enough to be clearly identified in the CCF maps (Figs.~\ref{aligned-planetary-trails_CARMENES} and \ref{aligned-planetary-trails_HARPS}).
Since we only used the sections of planetary trail that are located outside the region dominated by stellar iron lines (c.f. the CCF map in Fig.~\ref{CCF-map_CARMENES_Fe}), the detected signal is not affected by the stellar line residuals from the pulsations. Moreover, we computed the S/N maps of Fe by using the non-inverted $T$-$p$ profile from \cite{Nugroho2017} and observed negative S/N peaks with orbital parameters that are consistent with those found by using the inverted atmospheric profile (Fig.~\ref{SN_maps_Fe_un-inverted}).

Our results confirm the recent report of neutral iron and an atmospheric inversion layer in the dayside of WASP-33b \citep{Nugroho2020b}. As the model spectrum used for cross-correlation is based on an inverted $T$-$p$ profile, the detection of Fe is an unambiguous proof of a temperature inversion in the planetary atmosphere. The negative detection peaks obtained with a non-inverted $T$-$p$ profile further substantiate the existence of a thermal inversion layer. The obtained $K_\mathrm{p}$ values are close to the expected $K_\mathrm{p}$ ($231\pm3$\,km\,s$^{-1}$), which was calculated using the planetary orbital parameters. The strong detection of Fe strengthens the hypothesis that the species is significantly contributing to the heating of the upper planetary atmosphere \citep{Lothringer2018, Lothringer2019}. However, additional atomic and molecular species (e.g., Ti, Mg, AlO, SiO, CaO, FeH) and ions (e.g., \ion{Fe}{ii}, \ion{Mg}{ii}) may also contribute to maintain the atmospheric temperature inversion \citep{Lothringer2018, Lothringer2019, Gandhi2019}.

\begin{figure*}
	\centering
	\includegraphics[width=\textwidth]{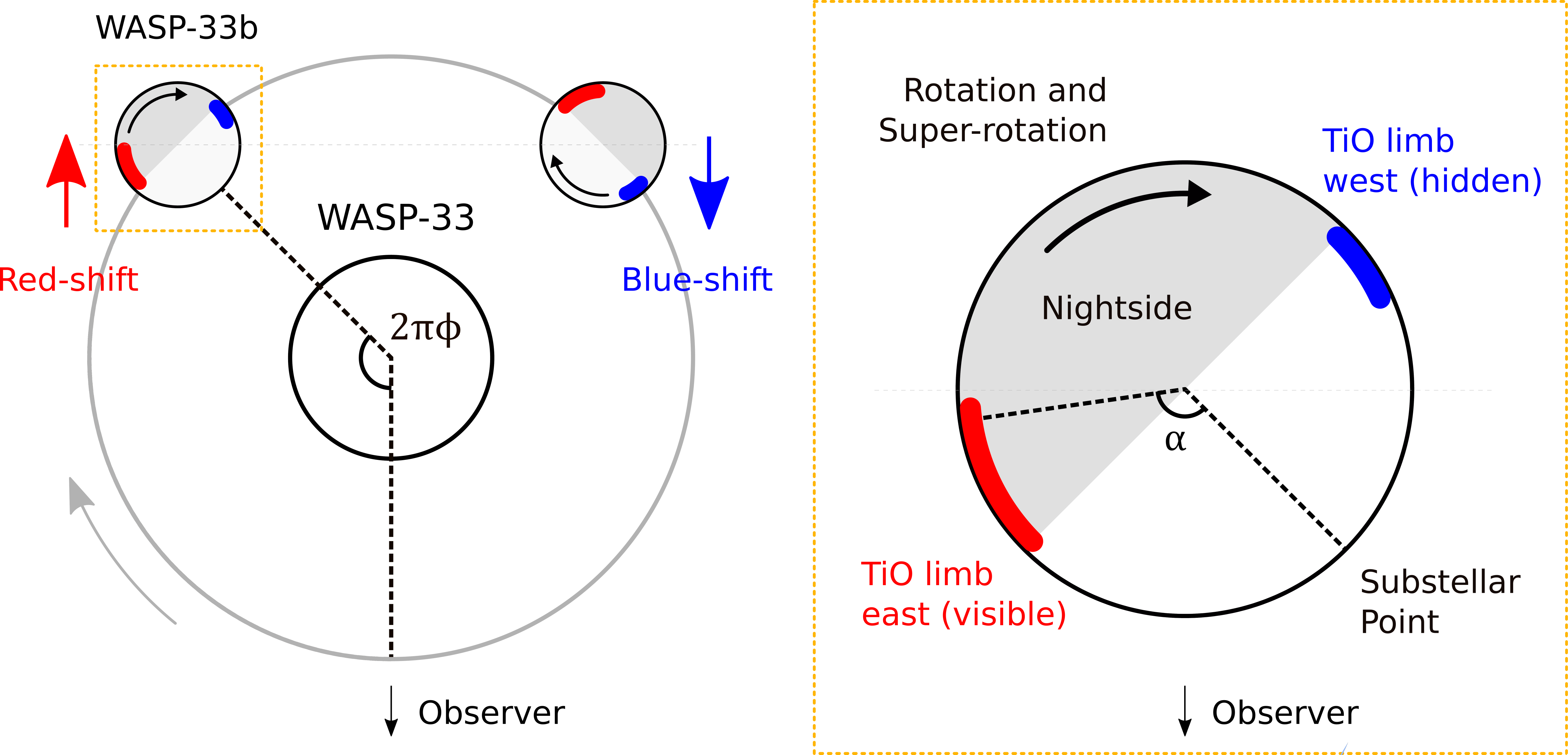}
	\caption{Illustration of the toy model. The left panel represents the WASP-33 system; the right panel illustrates a zoom in on the planet. The curved black arrows indicate the rotation direction and the gray arrow indicates the orbital motion direction. We assumed the presence of TiO near the terminator (indicated as red and blue regions). Due to the planetary rotation, the TiO signature is red-shifted before eclipse (red arrow) and blue-shifted after eclipse (blue arrow). The orbital phase is $\phi$ and the geographical longitude is denoted with $\alpha$ (west limb at --90\,$^\circ$; east limb at +90\,$^\circ$; substellar point at 0\,$^\circ$).}
	\label{Toy_Model_Setup}
\end{figure*}

\begin{figure*}
	\centering
	\includegraphics[width=\textwidth]{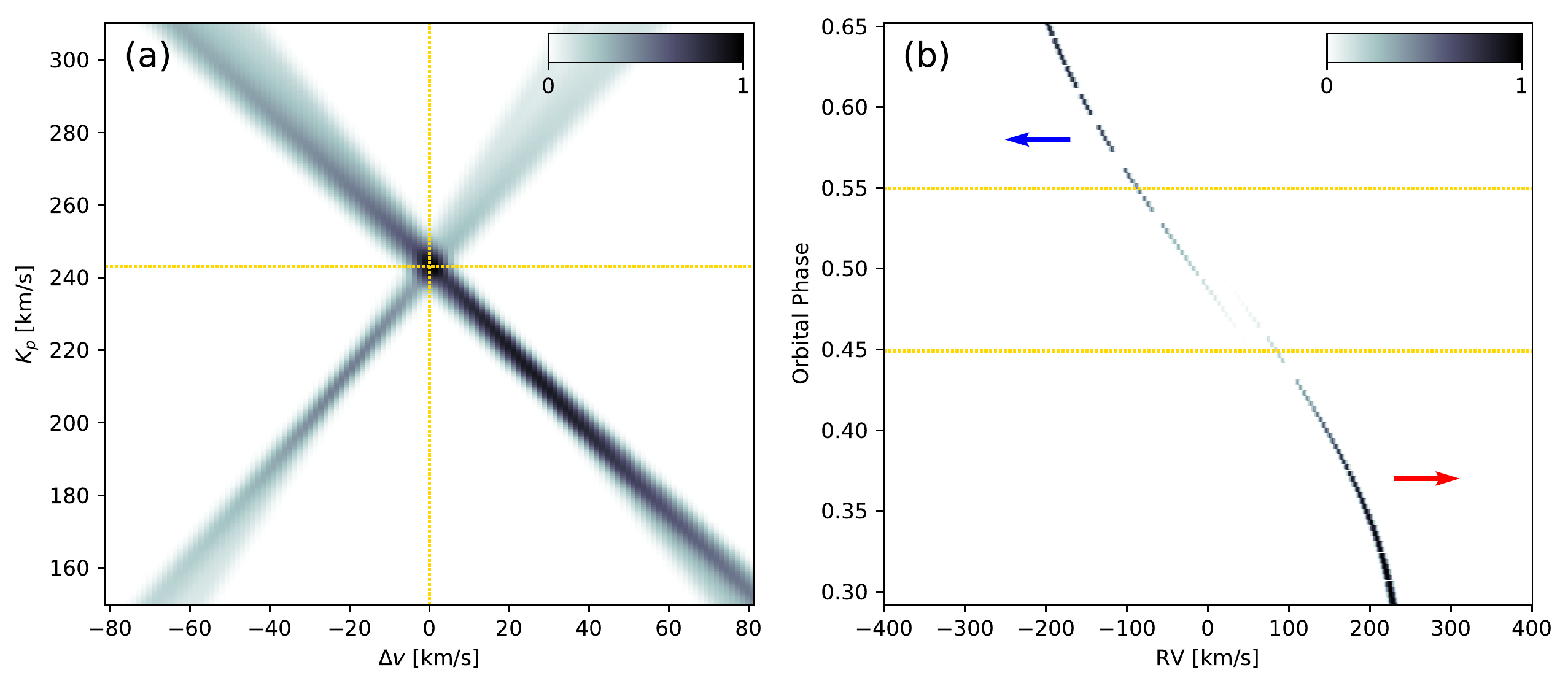}
	\caption{Solution of the toy model for the TiO signal. \textit{(a)} Modeled S/N map. The yellow lines indicate the peak value at $K_\mathrm{p}$\,=\,243.0\,km\,s$^{-1}$. \textit{(b)} Modeled CCF map in the stellar rest frame. The yellow lines indicate the begin and end of the secondary eclipse. We also show the continuation of the CCF trail during eclipse for a better understanding. Prior to the secondary eclipse only the terminator region at the east limb is visible to the observer, leading to a red-shift of the TiO signature (shift indicated by red arrow). After eclipse only the blue-shifted TiO signal from the terminator at the west limb is visible (shift indicated by blue arrow).}
	\label{SN_map_and_trail_toy-model}
\end{figure*}

\subsection{Evidence for TiO}
\label{Evidence for TiO}
We find evidence for TiO in all ten S/N maps of CARMENES, with each map corresponding to a different number of \texttt{SYSREM} iterations (see Fig.~\ref{SN_maps_CARMENES} left panel). In contrast, no significant signature at physically realistic values of $K_\mathrm{p}$ was detectable with HARPS-N. This non-detection is not surprising because the expected TiO emission feature is relatively weak in the HARPS-N wavelength range (Fig.~\ref{T-P-profiles-and-template-spectra}).

The TiO signal found with CARMENES peaks after six consecutive iterations and when we excluded an orbital phase interval around secondary eclipse (i.e., between 0.37 and 0.60, which is masked in Fig.~\ref{aligned-planetary-trails_CARMENES}). This suggests that inside this orbital phase range the planetary signal is not detectable. A map of the detection strength when excluding different orbital phase ranges is provided in Fig.~\ref{phase-exclusion-mapping_TiO}. We found the maximum S/N at $K_\mathrm{p}$\,=\,$248.0_{-2.5}^{+2.0}$\,km\,s$^{-1}$ and $\Delta \varv$\,=\,$0.0_{-2.6}^{+2.6}$\,km\,s$^{-1}$ with a significance level of S/N = 4.9 (cf. Table \ref{tab-results}). The measured evidence for TiO is in line with prior detections \citep{Haynes2015, Nugroho2017} and consistent with the systemic velocities calculated by \cite{Nugroho2017, Nugroho2020b}. However, $K_\mathrm{p}$ is located +17\,km\,s$^{-1}$ off from the expected value (231\,km\,s$^{-1}$). In comparison, \cite{Nugroho2017} also found a deviation of about +8\,km\,s$^{-1}$ from the expected $K_\mathrm{p}$ value.

Our detected signature of TiO is unlikely to originate from the host star. WASP-33 is an A-type star and hence, we expect the absence of significant stellar TiO concentrations (e.g., in stellar spots). Moreover, the planetary trail of TiO in Fig.~\ref{aligned-planetary-trails_CARMENES} is located outside the RV range that could potentially be affected by residual stellar TiO lines (i.e., outside $\pm \varv_\mathrm{rot}\sin i_*$ in the stellar rest frame).

\subsection{TiO depleted hot spot region}
\label{Discussion}
The TiO signal is located at a $K_\mathrm{p}$ value deviating from the expected one, while the Fe signal is consistent with the expected value.
To explain the $K_\mathrm{p}$ deviation, we propose the presence of a TiO-depleted hot spot region in the dayside atmosphere of WASP-33b. General circulation models predict a confined region with enhanced temperature, namely the hot spot, as a general feature in UHJ atmospheres \citep[e.g.,][]{Komacek2017, Parmentier2018, Arcangeli2019}. With an average dayside temperature of $T_\mathrm{day}$\,$\sim$\,3000\,K \citep{Zhang2018, Essen2020}, TiO is expected to be largely dissociated in the hot spot region. Consequently, the molecule will be more concentrated toward the terminator region of the planet. Moreover, theoretical studies predict Doppler shifts of hot Jupiter atmospheres due to global-scale winds with predominant super-rotation \citep[e.g.,][]{Showman2013, Zhang2017, Flowers2019, Beltz2020}. Therefore, we propose that the combination of a TiO-depleted hot spot region, the planetary rotation, and atmospheric winds could cause the observed deviation of $K_\mathrm{p}$.

To check the plausibility of this scenario, we implemented a simple toy model to compute the synthetic CCF map and the corresponding S/N map. In this toy model, we assumed that TiO is absent in the hot spot region and is only present in the regions close to the equator at both sides of the planetary limb. An example of the model setup is shown in Fig.~\ref{Toy_Model_Setup}. We divided the TiO regions into individual longitudinal grid points and assumed a Doppler shifted Gaussian profile as the CCF of each point. The velocity of each grid point is determined by the planetary rotation, the super-rotation of the atmosphere, and the planetary orbital motion. The velocity at the equator due to rotation was set to $\varv_\mathrm{rot}$\,=\,7\,km\,s$^{-1}$, which corresponds to a tidally locked planet.

To construct a model S/N map resembling that of the CARMENES observations, we defined a set of different combinations of the following parameters: geographical longitudes where TiO is present, velocities of the super-rotation and $K_\mathrm{p}$ values in the uncertainty range ($231\pm3$\,km\,s$^{-1}$). We simulated the S/N map for each combination and assessed the locations of the peak values. As a result, we found several parameter configurations that led to a peak in the model S/N map at $K_\mathrm{p}$ values between 240\,km\,s$^{-1}$ and 250\,km\,s$^{-1}$. To produce a S/N peak in this range, our model tends to favor the presence of TiO on the night hemisphere or close to the terminator regions. However, the nightside atmosphere has a significantly lower average temperature than the dayside \citep{Essen2020}. Hence, the nightside atmosphere is unlikely to possess a temperature inversion and we do not expect that the TiO emission originates from the nightside hemisphere. Instead, we hypothesize that the TiO emission signals are from the regions close to the planetary terminators. We also suggest that a super-rotating atmosphere may transport heated gas across the eastward terminator, leading to a significant presence of TiO in a restricted region of the nightside. Figure~\ref{SN_map_and_trail_toy-model} shows a toy model solution for this scenario, adopting a super-rotating atmosphere with $\varv_\mathrm{wind}$\,=\,8\,km\,s$^{-1}$ and two TiO regions at longitudes 90\,$^\circ$\, to \,130\,$^\circ$ (east limb) and --70\,$^\circ$\, to --90\,$^\circ$ (west limb) (Fig.~\ref{Toy_Model_Setup}). This led to a maximum S/N located at $K_\mathrm{p}$\,=\,243.0\,km\,s$^{-1}$, which is close to the orbital parameters of the planet obtained with CARMENES from the TiO signal. In order to obtain a similar $K_\mathrm{p}$ as the CARMENES result, only one TiO region must be visible at a given phase, otherwise we would observe a double-peak CCF. Figure~\ref{Toy_Model_Setup} shows that the region at the east planetary limb is only visible before the eclipse while the west limb region is only visible after the eclipse. Although the toy model can explain the observed TiO signal, we emphasize that the model is not constrained sufficiently well to retrieve the actual parameters of the global circulation and the TiO distribution. Due to the simplicity of the toy model, we also refrain from giving the uncertainties of the parameters. The parameters of global circulation could probably be retrieved from a more comprehensive model of the atmosphere along with observations with higher S/N.

Our toy model also predicts that the TiO signal weakens at orbital phases close to the eclipse as the TiO-depleted hot spot faces toward the observer. This prediction is consistent with the observed results. We found the strongest TiO peak in the S/N map when excluding the orbital phases between 0.37 and 0.60 (Fig.~\ref{phase-exclusion-mapping_TiO}). This indicates that the spectra inside this phase range probably carry a very weak TiO signal that is below the noise level.
On the other hand, Fe is not depleted in the hot spot region but distributed more homogeneously over the planetary dayside. 
We emphasize that the suggested scenario of a TiO-free hot spot is consistent with our choice of using two different $T$-$p$ profiles. We assumed atmospheric profiles with a higher temperature for Fe and a moderate temperature for TiO. Each profile may describe the average thermal conditions for the specific species. To verify this hypothesis, we reassessed the detection strengths by exchanging the two temperature profiles used to calculate the TiO and Fe model spectra. The coordinates of the S/N peaks do not change significantly (cf. Fig.~\ref{SN_maps_switch_Nugroho-vs-W189}). This indicates that the detection peaks are mainly caused by the presence of a thermal inversion layer. However, the detection strengths are lower than the results in Sections \ref{Detection of Fe} and \ref{Evidence for TiO}, which is an additional hint toward TiO emission at moderate thermal conditions and Fe emission at higher temperatures.

\subsection{Comparison of line profiles}

We also assessed the width of the detected significance peaks by fitting a Gaussian function to the signals (Fig.~\ref{CCF_width_comparison}). We found FWHM values of the Fe signal equal to \mbox{$8.6\pm1.0$\,km\,s$^{-1}$} and $6.7\pm0.8$\,km\,s$^{-1}$ for CARMENES and HARPS-N, respectively. The FWHM of the TiO signal is equal to \mbox{$4.0\pm0.7$\,km\,s$^{-1}$}. The results are summarized in Table \ref{tab-results}.

The CCF of Fe is significantly broader than that of TiO (cf. Figs.~\ref{SN_maps_CARMENES} and \ref{CCF_width_comparison} for comparison; FWHM values in Table \ref{tab-results}). This indicates that the line width of Fe is larger when compared to the TiO lines. Therefore, we suggest that the two chemical species experience a different level of rotational broadening, with Fe being affected more strongly than TiO. The narrow TiO line profile can be explained by the presence of the TiO-depleted hot spot, which produces a localized TiO concentration and a less rotational-broadened line profile.
We checked this assumption by simulating two cross-correlation functions: the auto-correlation of the non-broadened TiO model spectrum; the cross-correlation between the non-broadened Fe model and a rotationally broadened Fe model, assuming a tidally locked rotational velocity (i.e., 7\,km\,s$^{-1}$ at the equator). The widths of the simulated CCFs are close to the widths of the observed CCFs (Fig.~\ref{autocorrelation-with-broadened-template}), indicating that the profile of the Fe lines is more strongly rotationally broadened than the profile of the TiO lines.

\begin{figure*}
	\centering
	\includegraphics[width=\textwidth]{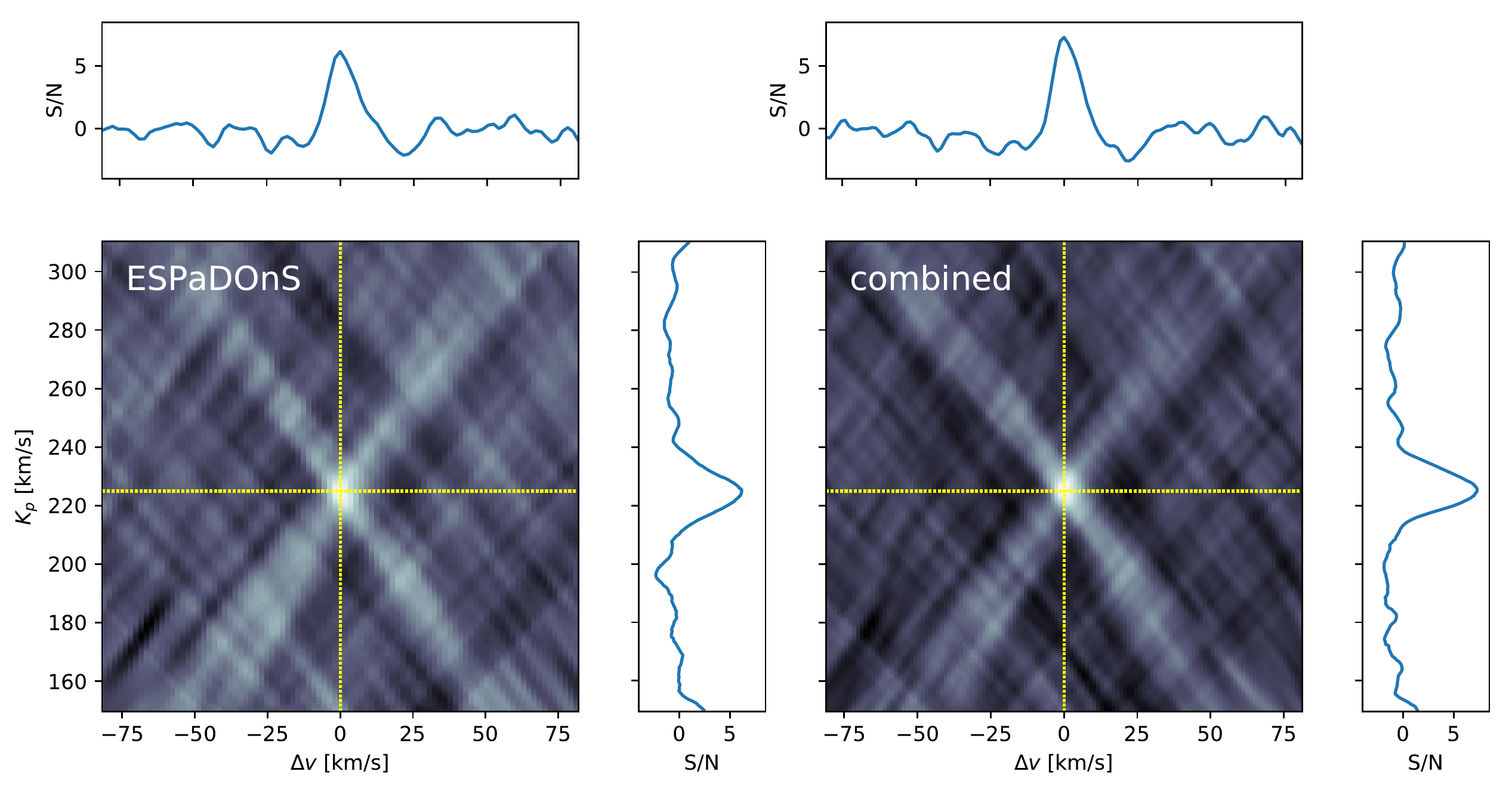}
	\caption{S/N maps of Fe for ESPaDOnS (left panel) and a combination of all instruments (CARMENES, HARPS-N and ESPaDOnS; right panel). The detection significance peaks in both cases at $K_\mathrm{p}$\,=\,$225.0$\,km\,s$^{-1}$ and $\Delta \varv$\,=\,$0.0$\,km\,s$^{-1}$ with a S/N value of 6.2 and 7.3, respectively. The peak coordinates in the S/N maps are indicated by the yellow dashed lines; cross-sections of the S/N peaks are reported in the horizontal and vertical panels.}
	\label{SN_map_Fe_ESPADONS-and-combined}
\end{figure*}

\subsection{Comparison with previous work}
\label{Comparison with previous work}
In contrast to the results of this work and previous studies \citep{Haynes2015, Nugroho2017}, a recent investigation by \cite{Herman2020} does not report a significant detection of TiO in the atmosphere of WASP-33b. Their measurements with ESPaDOnS at the Canada-France-Hawaii telescope (CFHT) are comparable to our data, as the authors use a cumulative exposure time equivalent to $\sim$\,1.4 times of our observation with CARMENES. At $K_\mathrm{p}$\,$\sim$\,250\,km\,s$^{-1}$ and $\varv_\mathrm{sys}$\,$\sim$\,0\,km\,s$^{-1}$ in their Fig.~4 a weak pattern is visible, which resembles the peak region in the S/N map from our work (cf. left panel in Fig.~\ref{SN_maps_CARMENES}). The orientation of the weak pattern (from upper left to lower right in their Fig.~4) indicates that most of the contribution to this feature is from the observations before the eclipse.

The authors use the Plez'12 \citep{Plez2012} database to compute the TiO model spectrum instead of the more precise \texttt{ToTo} \texttt{ExoMol} \citep{McKemmish2019}. Thus, we analyzed the data from \cite{Herman2020} to investigate whether the weak signature at $K_\mathrm{p}$\,$\sim$\,250\,km\,s$^{-1}$ is caused by a real TiO signal or by line list dependent effects \citep{Merritt2020}. We downloaded the spectra from the CFHT archive and applied the same data reduction procedures as for the CARMENES and HARPS-N spectra. We used the \texttt{ToTo} \texttt{ExoMol} TiO model spectrum (as in Fig.~\ref{T-P-profiles-and-template-spectra}) to perform the cross-correlation. Our re-analyzed ESPaDOnS result shows a weak TiO signature with S/N\,$\sim$\,3, which is similar to the weak pattern in Fig.~4 of \cite{Herman2020}. The obtained S/N map of TiO is presented in Fig.~\ref{SN_map_TiO_prior-to-eclipse_ESPADONS}. This weak feature is located close to our TiO detection with CARMENES. Therefore, this ESPaDOnS feature could be a real signature of TiO. However, the TiO signature in the ESPaDOnS spectra appears to be absent after the eclipse in contrast to the CARMENES observation. This could be due to the existence of temporal variability in global circulation \citep{Komacek2020}.

Very recently, \cite{Serindag2021} reassessed the TiO observation from \cite{Nugroho2017}, but did not confirm the results from this previous study. In line with \cite{Serindag2021}, we used the same line list \texttt{ToTo} \texttt{ExoMol} and adopted a similar approach by assuming a common optimal number of \texttt{SYSREM} iterations. They find a significant TiO detection with a $K_\mathrm{p}$ that is similar to our value, but different from the value published by \cite{Nugroho2017}. In contrast, we identified the molecular signature at a $\Delta \varv$ that is consistent with the planetary rest frame. We assumed the presence of a TiO-depleted hot spot to explain the $K_\mathrm{p}$ deviation of the detection peak, a hypothesis that has to be confirmed in future studies. Otherwise, we cannot exclude the presence of a spurious signal. This aspect is also discussed by \cite{Serindag2021}, who suggest that previous claims of TiO may be false positives. The contradicting results of different studies on TiO in the atmosphere of WASP-33b \citep{Haynes2015, Nugroho2017, Herman2020} highlight the critical role of accurate line lists and the potential need of different data reduction techniques and spectral models in further investigations.

\subsection{Additional analysis of the Fe signal}
\label{Combining the Fe signal}
In addition to TiO, we also searched for Fe features in the ESPaDOnS data. We applied the same reduction procedures as described in Sections \ref{Data reduction} and \ref{Methods}. The velocity steps were set to 1.8\,km\,s$^{-1}$, which corresponds to the mean pixel spacing of the instrument. The signature of Fe is clearly detected with a peak S/N of 6.2 after nine \texttt{SYSREM} iterations (Fig.~\ref{SN_map_Fe_ESPADONS-and-combined}). The signal is located at $K_\mathrm{p}$\,=\,$225.0_{-5.0}^{+3.5}$\,km\,s$^{-1}$ and $\Delta \varv$\,=\,$0.0_{-3.6}^{+3.6}$\,km\,s$^{-1}$, which is in agreement with the results from CARMENES and HARPS-N. The FWHM of the peak CCF is measured as $11.1\pm1.1$\,km\,s$^{-1}$.

We further calculated a combined S/N map of Fe using the data from all three instruments (CARMENES, HARPS-N, ESPaDOns). For this purpose, we used a common velocity step equal to 1.3\,km\,s$^{-1}$. We firstly merged the CCFs of all the spectra and then computed the S/N map following the description in Section \ref{Searching for atmospheric features}. The final combined S/N map is presented in the right panel of Fig.~\ref{SN_map_Fe_ESPADONS-and-combined}. The resulting detection peak with S/N\,=\,7.3 is located at $K_\mathrm{p}$\,=\,$225.0_{-3.5}^{+4.0}$\,km\,s$^{-1}$ and $\Delta \varv$\,=\,$0.0_{-2.6}^{+2.6}$\,km\,s$^{-1}$. The measured FWHM of the combined Fe peak CCF is $9.2\pm0.7$\,km\,s$^{-1}$.

\begin{table}
	\caption{Summary of results.} 
	\label{tab-results} 
	\centering     
	\renewcommand{\arraystretch}{1.2} 
	\begin{threeparttable}
		\begin{tabular}{l c c c c r}      
			\hline\hline  
			\noalign{\smallskip}
			Instrument & S/N  & $K_\mathrm{p}$  & $\Delta \varv$  & FWHM      \\           
			&      & [km\,s$^{-1}$]  & [km\,s$^{-1}$]  & [km\,s$^{-1}$] \\
			\noalign{\smallskip}
			\hline    
			\noalign{\smallskip}
			\textit{TiO} & & & & \\
			\noalign{\smallskip}
			CARMENES    & 4.9    & $248.0_{-2.5}^{+2.0}$  & $0.0_{-2.6}^{+2.6}$ & $4.0\pm0.7$ \\ 
			HARPS-N     & \multicolumn{4}{c}{no significant detection}           \\
			ESPaDOnS    & \multicolumn{4}{c}{no significant detection}           \\		
			\noalign{\smallskip}
			\hline
			\noalign{\smallskip}
			\textit{Fe} & & & & \\
			\noalign{\smallskip}
			CARMENES    & 5.7    & $228.0_{-5.0}^{+3.5}$  & $1.3_{-3.9}^{+3.9}$  & $8.6\pm1.0$ \\	
			HARPS-N     & 4.5    & $225.0_{-5.0}^{+2.0}$  & $-0.8_{-2.4}^{+4.0}$ & $6.7\pm0.8$ \\	
			ESPaDOnS    & 6.2    & $225.0_{-5.0}^{+3.5}$  & $0.0_{-3.6}^{+3.6}$  & $11.1\pm1.1$ \\				
			Combined    & 7.3    & $225.0_{-3.5}^{+4.0}$  & $0.0_{-2.6}^{+2.6}$  & $9.2\pm0.7$ \\		
			\noalign{\smallskip}
			\hline                               
		\end{tabular}
	\end{threeparttable}      
\end{table}

%

\section{Conclusions}
\label{Conclusions}
We observed the dayside of WASP-33b at high spectral resolution with the CARMENES and HARPS-N spectrographs. By using the cross-correlation technique, we detected Fe and found strong evidence for the presence of TiO. Both species show emission spectra, which confirms the presence of a temperature inversion claimed by prior studies \citep{Haynes2015, Nugroho2017, Nugroho2020b}.
For TiO, we found the signal peak at $K_\mathrm{p}$\,=\,$248.0_{-2.5}^{+2.0}$\,km\,s$^{-1}$, which deviates from the literature value by +17\,km\,s$^{-1}$ \citep{Kovacs2013, Lehmann2015, Nugroho2020b}. In contrast, we detected Fe at $K_\mathrm{p}$\,=\,$225.0_{-3.5}^{+4.0}$\,km\,s$^{-1}$, which is consistent with the literature values. The observed CCF of Fe is broader than that of TiO, indicating that the Fe lines are broader than the TiO lines.

We hypothesize that a TiO-depleted hot spot is present in the atmosphere of WASP-33b. Since TiO is suggested to be thermally dissociated in the hot spot region, we suppose that the observed TiO signal originates from locations close to the terminators. Our toy model suggests that this could lead to the observed deviation of $K_\mathrm{p}$ from the literature value. Such a scenario is also supported by the observed narrow line profile of TiO. Because TiO may be restricted to regions outside the hot spot while a homogeneous Fe distribution is expected in the dayside hemisphere, the TiO lines are less broadened by the planetary rotation compared to the Fe lines.

Although temperature inversions have been detected in a number of UHJs, the underlying formation mechanisms are still a matter of discussion. Our results suggest that atomic species and metal oxides are both involved in the heating mechanism, which is required to maintain a thermal inversion layer. Observations with higher S/N and the inclusion of 3D atmospheric structure into spectral modeling will be beneficial for further advances in exploring the atmospheres of extremely irradiated planets. 

\begin{acknowledgements}
	
	CARMENES is an instrument at the Centro Astron\'omico Hispano-Alem\'an (CAHA) at Calar Alto (Almer\'{\i}a, Spain), operated jointly by the Junta de Andaluc\'ia and the Instituto de Astrof\'isica de Andaluc\'ia (CSIC).
	
	CARMENES was funded by the Max-Planck-Gesellschaft (MPG), 
	the Consejo Superior de Investigaciones Cient\'{\i}ficas (CSIC),
	the Ministerio de Econom\'ia y Competitividad (MINECO) and the European Regional Development Fund (ERDF) through projects FICTS-2011-02, ICTS-2017-07-CAHA-4, and CAHA16-CE-3978, 
	and the members of the CARMENES Consortium 
	(Max-Planck-Institut f\"ur Astronomie,
	Instituto de Astrof\'{\i}sica de Andaluc\'{\i}a,
	Landessternwarte K\"onigstuhl,
	Institut de Ci\`encies de l'Espai,
	Institut f\"ur Astrophysik G\"ottingen,
	Universidad Complutense de Madrid,
	Th\"uringer Landessternwarte Tautenburg,
	Instituto de Astrof\'{\i}sica de Canarias,
	Hamburger Sternwarte,
	Centro de Astrobiolog\'{\i}a and
	Centro Astron\'omico Hispano-Alem\'an), 
	with additional contributions by the MINECO, 
	the Deutsche Forschungsgemeinschaft through the Major Research Instrumentation Programme and Research Unit FOR2544 ``Blue Planets around Red Stars'', 
	the Klaus Tschira Stiftung, 
	the states of Baden-W\"urttemberg and Niedersachsen, 
	and by the Junta de Andaluc\'{\i}a.
	
	We acknowledge financial support from the
	Deutsche Forschungsgemeinschaft through the priority program SPP 1992 ``Exploring the Diversity of Extrasolar Planets'' (RE 1664/16-1), the Research Unit FOR2544 ``Blue Planets around Red Stars'' (RE 1664/21-1), and grant~CA 1795/3,
	the Agencia Estatal de Investigaci\'on of the Ministerio de Ciencia, Innovaci\'on y Universidades and the ERDF through projects
	PID2019-109522GB-C5[1:4]/AEI/10.13039/501100011033,	
	PID2019-110689RB-I00/AEI/10.13039/501100011033,  
	and the Centre of Excellence ``Severo Ochoa'' and ``Mar\'ia de Maeztu'' awards to the Instituto de Astrof\'isica de Canarias (SEV-2015-0548), Instituto de Astrof\'isica de Andaluc\'ia (SEV-2017-0709), and Centro de Astrobiolog\'ia (MDM-2017-0737), 
	the European Research Council under the European Union's Horizon 2020 research and innovation program (832428),
	and the Generalitat de Catalunya/CERCA programme.

	This work is based on observations made with the Italian Telescopio Nazionale Galileo (TNG) operated on the island of La Palma by the Fundación Galileo Galilei of the INAF (Istituto Nazionale di Astrofisica) at the Spanish Observatorio del Roque de los Muchachos of the Instituto de Astrof\'isica de Canarias.
	
\end{acknowledgements}

\bibliographystyle{aa} 

\bibliography{W33-TiO-Fe-refer}

\appendix

\section{Validation of the TiO line list}
\label{Validation of the TiO line list}	
We assessed the quality of the \texttt{ToTo} \texttt{ExoMol} line list via comparison of the line positions with the spectrum of Barnard's star \citep{Reiners2018} that is dominated by TiO absorption features. The TiO transmission model spectrum was computed with \texttt{petitRADTRANS} by using the VMR and $T$-$p$ profile from Section \ref{Spectral models}. The analysis covered both the CARMENES VIS channel and HARPS-N wavelength ranges. We convolved the transmission model spectrum with the instrument profiles and removed large-scale features with a Gaussian high-pass filter (25\,pixels for CARMENES; 75\,pixels for HARPS-N). The filter was also applied to the high-resolution spectrum of Barnard's star. We then computed the CCF between the spectrum and the transmission model. The considered Doppler shifts relative to the stellar rest frame were between --80.6\,km\,s$^{-1}$ and +80.6\,km\,s$^{-1}$ in steps of 1.3\,km\,s$^{-1}$ for CARMENES. We applied Doppler shifts between --80.0\,km\,s$^{-1}$ and +80.0\,km\,s$^{-1}$ in steps of 0.8\,km\,s$^{-1}$ in the HARPS-N analysis. The line list was assumed to be accurate if a prominent CCF peak (>\,3$\sigma$) was present at the systemic velocity of Barnard's star.

\begin{figure*}
	\centering
	\includegraphics[width=\textwidth]{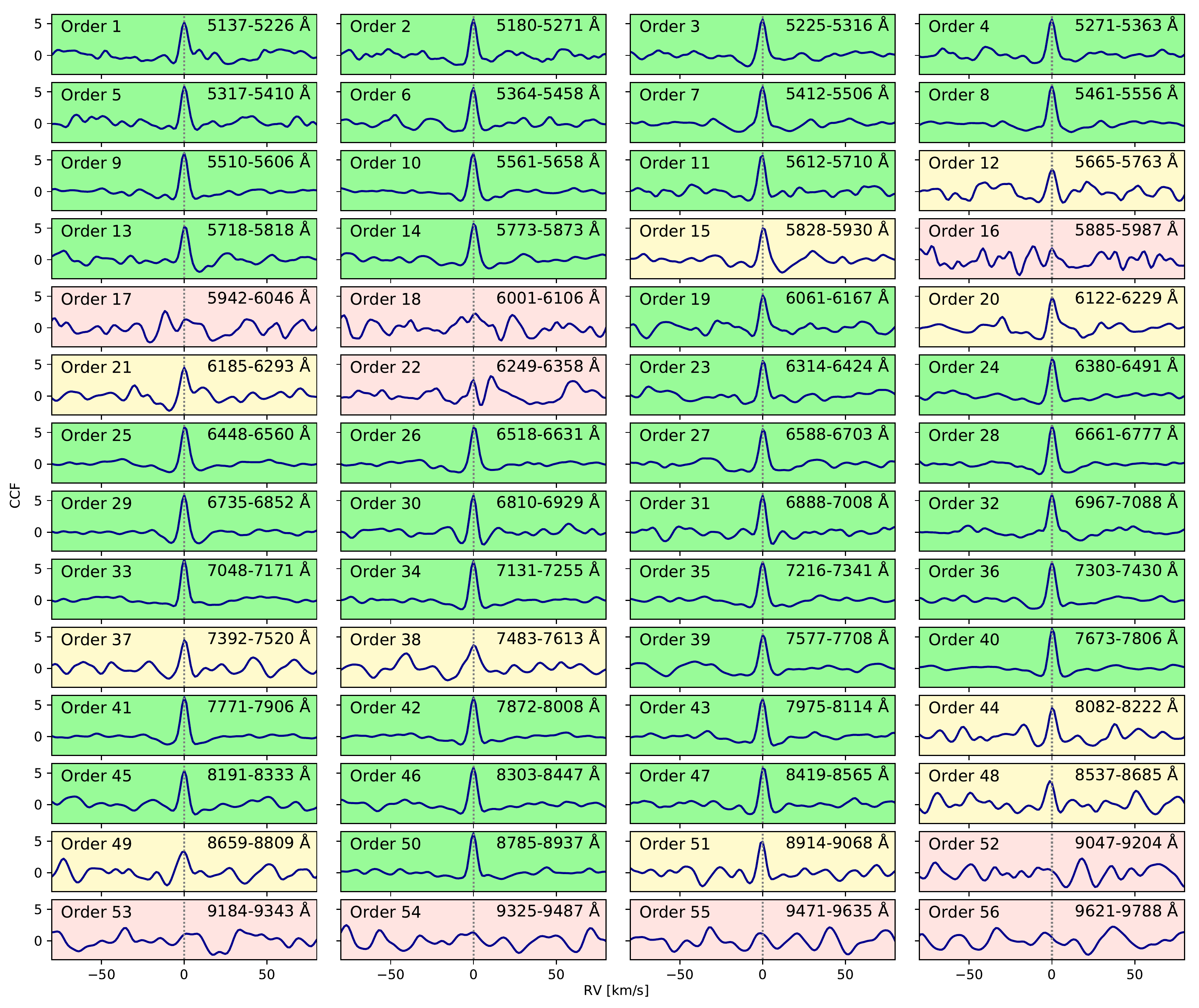}
	\caption{CCFs between the TiO transmission model and the high-resolution spectrum of Barnard's star for CARMENES. The x-axis represents the radial velocity offset from the stellar rest frame; the y-axis measures the CCF in units of standard deviation. The yellow and green shaded panels indicate the spectral orders with a CCF peak greater than three and five standard deviations, respectively. Red shaded panels represent spectral orders not showing any correlation.}
	\label{CCF-with-Barnards-star_CARMENES}
\end{figure*}

\begin{figure*}
	\centering
	\includegraphics[width=\textwidth]{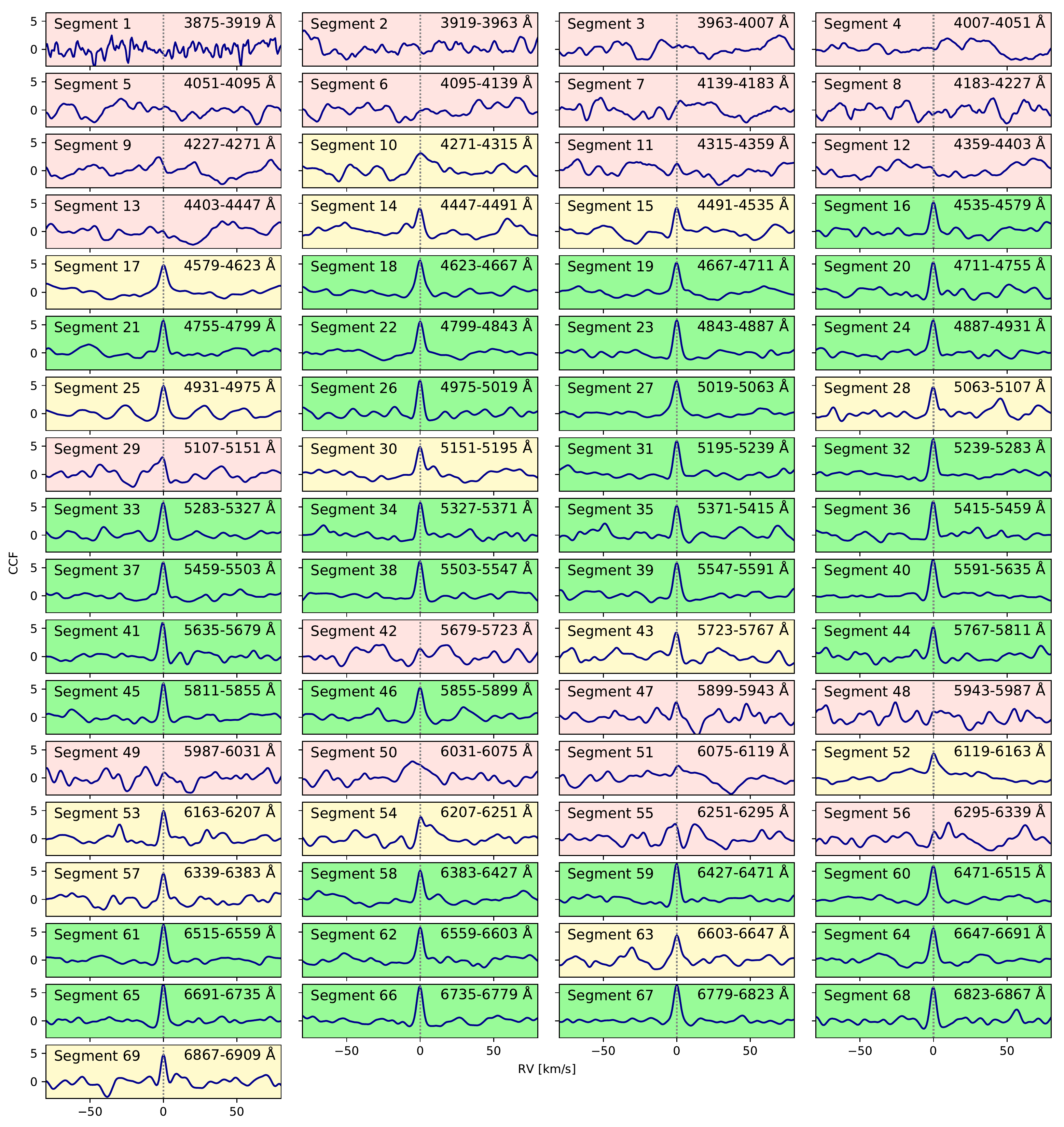}
	\caption{Same as Fig.~\ref{CCF-with-Barnards-star_CARMENES}, but for wavelength segments (cf. Section \ref{Pre-processing the spectra}) of HARPS-N.}
	\label{CCF-with-Barnards-star_HARPS}
\end{figure*}

We found prominent cross-correlation signals for most of the spectral orders, except for CARMENES spectral orders [16:18,~22,~52:56] and HARPS-N spectral orders (segments) [1:9,~11:13,~29,~42,~47:51,~55,~56]. This is consistent with the line list analysis from \cite{McKemmish2019} and shows the improved precision of \texttt{ToTo} \texttt{ExoMol} at wavelengths shorter than 6000\,\r{A} in comparison to the line list Plez'98 \citep{Plez1998, Hoeijmakers2015}. An order-wise plot of the CCF is given in Fig.~\ref{CCF-with-Barnards-star_CARMENES} and Fig.~\ref{CCF-with-Barnards-star_HARPS} for CARMENES and HARPS-N, respectively. Spectral orders at wavelengths with a poor line list were not considered in the TiO analysis.

\section{Injection-recovery test}
\label{Injection-recovery test}
After scaling the TiO model spectrum from Section \ref{Spectral models} by a factor of 3, we convolved it with the respective instrument profiles of CARMENES and HARPS-N. We then shifted the convolved model spectrum with the planetary orbital RV and injected the model spectrum into the raw spectra (1D spectra from the instrument pipelines). The raw spectra with the injected model were subsequently processed in the same way as described in Section~\ref{Data reduction} and cross-correlated with the convolved TiO model spectrum in Fig.~\ref{T-P-profiles-and-template-spectra}. This resulted in a cross-correlation matrix ($\overline{\mathrm{CCF}}_\mathrm{inj}$) for each observation and spectral order.

We computed a S/N map for all $\overline{\mathrm{CCF}}_\mathrm{inj}$s and identified the spectral orders that allowed us to recover the injected model spectrum (good orders). We assumed that only these spectral orders will give a contribution to the detection of the real planetary signature. In contrast, we assumed that spectral orders with no recovery of the injected model spectrum (bad orders) will not contribute to the detection of the real planetary signature. The following metric was applied to discriminate between good and bad orders. Good orders were supposed to detect the injected model spectrum (S/N > 3 at the injected $K_\mathrm{p}$ and $\Delta \varv$ position) at least for one specific number of \texttt{SYSREM} iterations in the S/N map. Bad orders were supposed to not show a significant detection of the injected model spectrum (S/N < 3).

\begin{figure*}
	\centering
	\includegraphics[width=\textwidth]{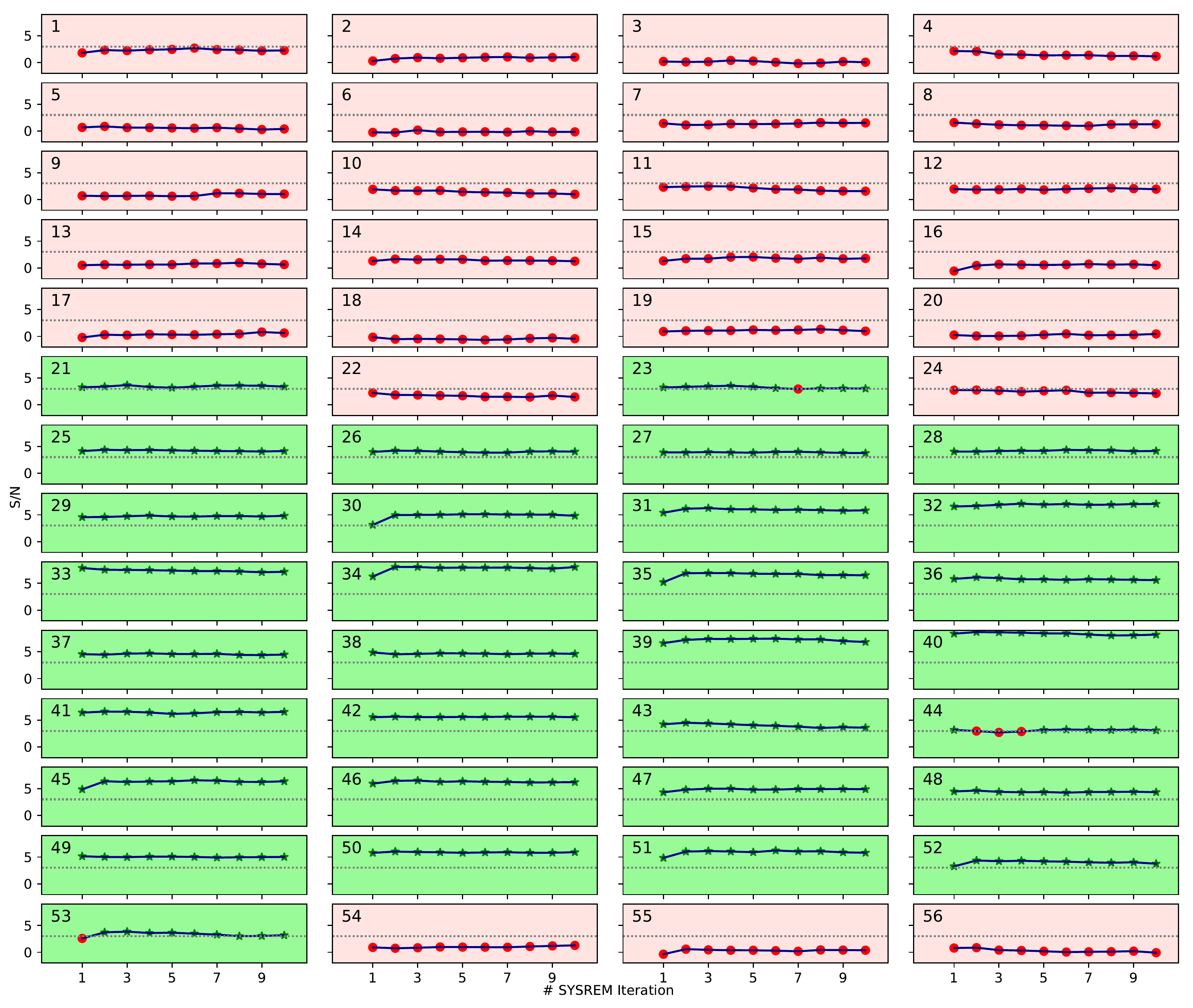}
	\caption{Strength of the injected TiO model spectrum. The x-axis represents the number of consecutive \texttt{SYSREM} iterations; the y-axis measures the S/N detection strength of the injected model spectrum. Each panel is labeled with the number of the corresponding spectral order. Data points marked with green stars (\textcolor{teal}{$\star$}) correspond to a detection, red circles (\textcolor{red}{$\bullet$}) to a non-detection of the injected model. The horizontal gray line corresponds to S/N = 3. We consider a spectral order to be good, if the injected model spectrum is detected at S/N > 3 for at least one specific number of \texttt{SYSREM} iterations. Spectral orders shaded with green are considered to be good orders, orders shaded with red to be bad orders. Although we got detections of the injected model in orders 50 and 51, we observed a strong enhancement of noise the in the final S/N detection map if we included them (cf. Section \ref{Searching for atmospheric features}). We concluded that these orders have an increased noise level and excluded them from the list of good orders.}
	\label{3x-injection_CARMENES}
\end{figure*}

\begin{figure*}
	\centering
	\includegraphics[width=\textwidth]{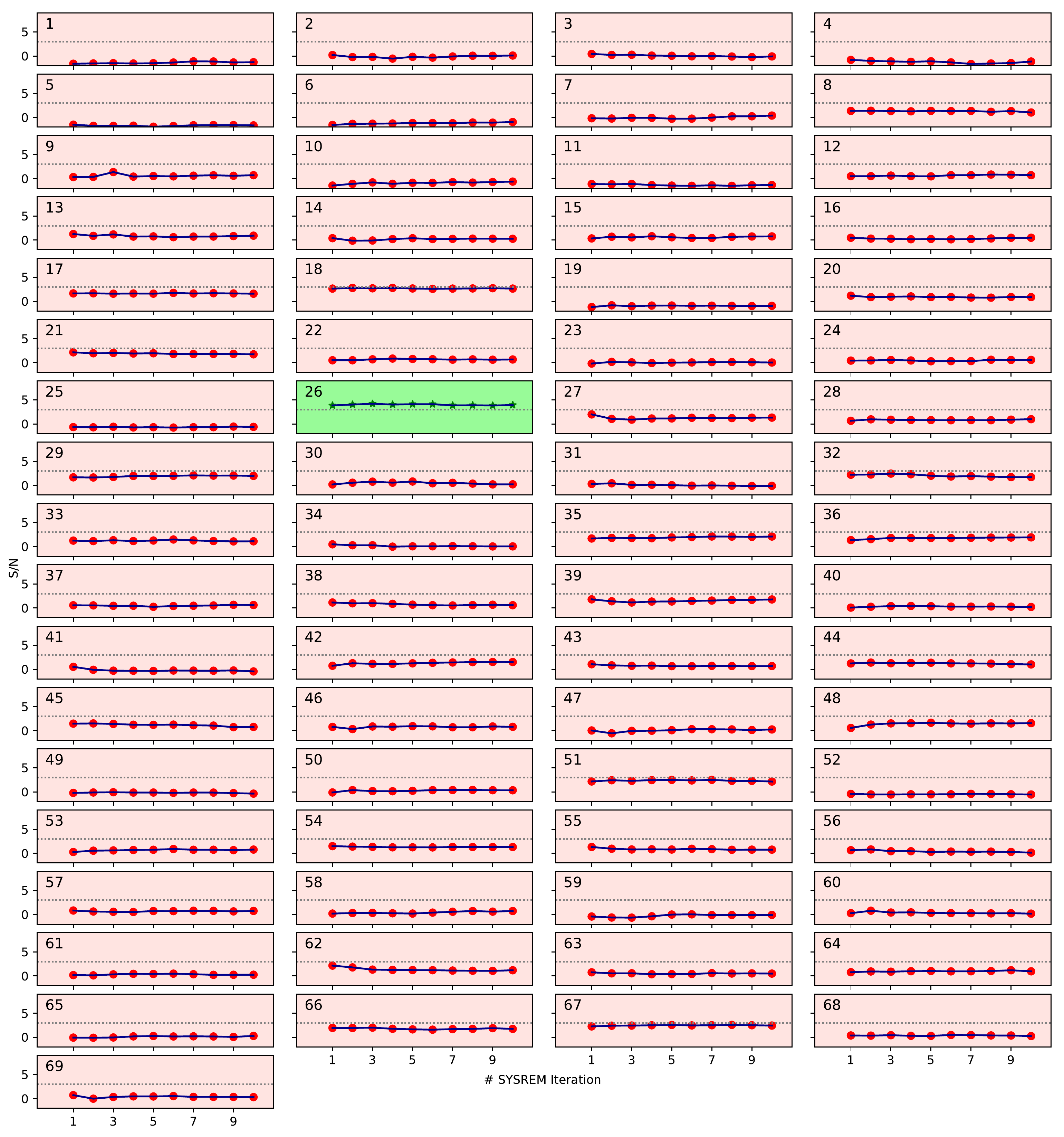}
	\caption{Same as Fig.~\ref{3x-injection_CARMENES}, but for wavelength segments (cf. Section \ref{Pre-processing the spectra}) of HARPS-N. We retrieved the injected model planet spectrum only in one segment.}
	\label{3x-injection_HARPS}
\end{figure*}

As a result, we found that the spectral orders number [1:20,~22,~24,~50,~51,~54:56] will not contribute to the detection of TiO in the CARMENES observation. For HARPS-N we recovered the injected model spectrum only in one order (segment) out of 69, that is order number 26. This suggests that the TiO signal will not be detectable with the HARPS-N data even if present in the data. The recovered detection strengths are shown for each spectral order in Fig.~\ref{3x-injection_CARMENES} and Fig.~\ref{3x-injection_HARPS}.

\section{Additional figures}

\begin{figure*}
	\centering
	\includegraphics[width=\textwidth]{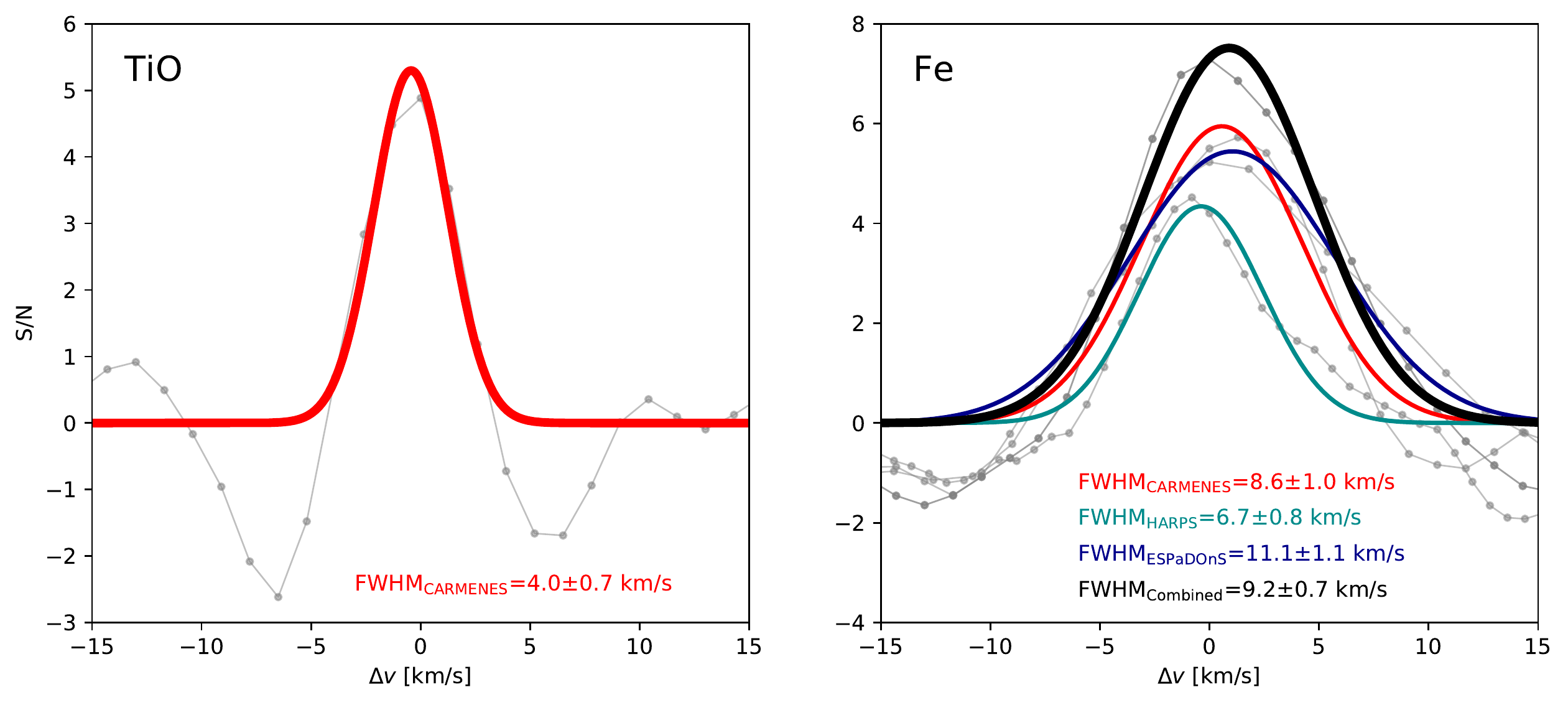}
	\caption{Comparison between the CCF widths of the detection peaks of TiO and Fe. The Fe signal is broader when compared to TiO. The observed CCFs are represented by the gray circles (\textcolor{gray}{$\bullet$}). The Gaussian fit functions are presented by the thick solid lines with different colors denoting different instruments. Red~(\textcolor{red}{\textbf{\textemdash}}):~CARMENES; green~(\textcolor{teal}{\textbf{\textemdash}}): HARPS-N; blue~(\textcolor{blue}{\textbf{\textemdash}}): ESPaDOnS; black~(\textcolor{black}{\textbf{\textemdash}}): Combined signal of CARMENES, HARPS-N, ESPaDOnS.}
	\label{CCF_width_comparison}
\end{figure*}

\begin{figure*}
	\centering
	\includegraphics[width=\textwidth]{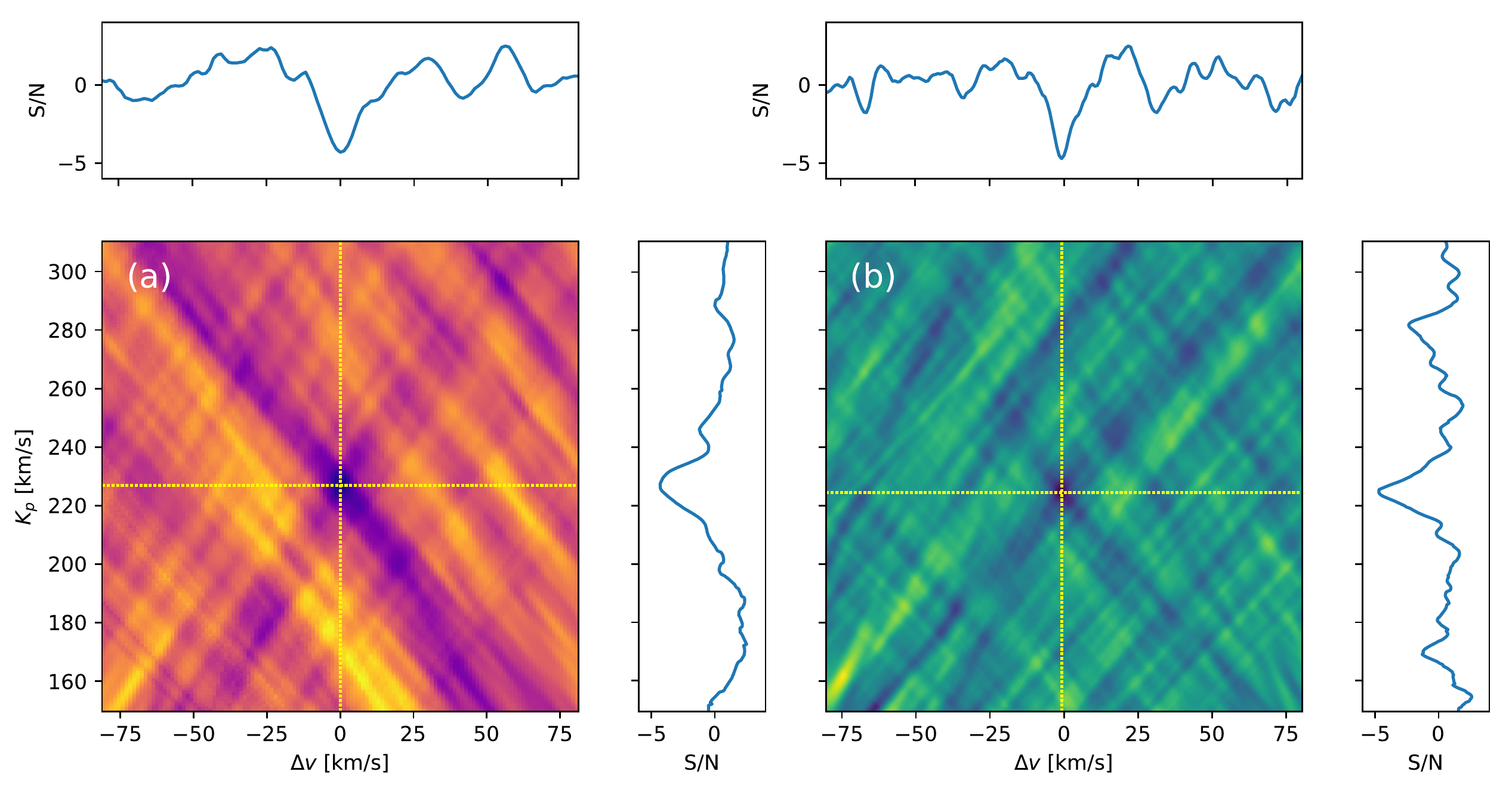}
	\caption{S/N maps of Fe obtained by using a non-inverted $T$-$p$ profile for cross-correlation \citep{Nugroho2017}. Panel \textit{(a)} shows the anticorrelation signal of Fe in the CARMENES data (S/N\,=\,-4.3); panel \textit{(b)} shows the anticorrelation with the HARPS-N data (S/N\,=\,-4.6). The coordinates of the negative S/N peaks are indicated by the yellow dashed lines. The cross-sections of the negative S/N peaks are reported in the horizontal and vertical panels. Each S/N map corresponds to the \texttt{SYSREM} iteration number that maximizes the detection strength in Sections \ref{Detection of Fe} and~\ref{Evidence for TiO}.}
	\label{SN_maps_Fe_un-inverted}
\end{figure*}

\begin{figure*}
	\centering
	\includegraphics[width=0.5\textwidth]{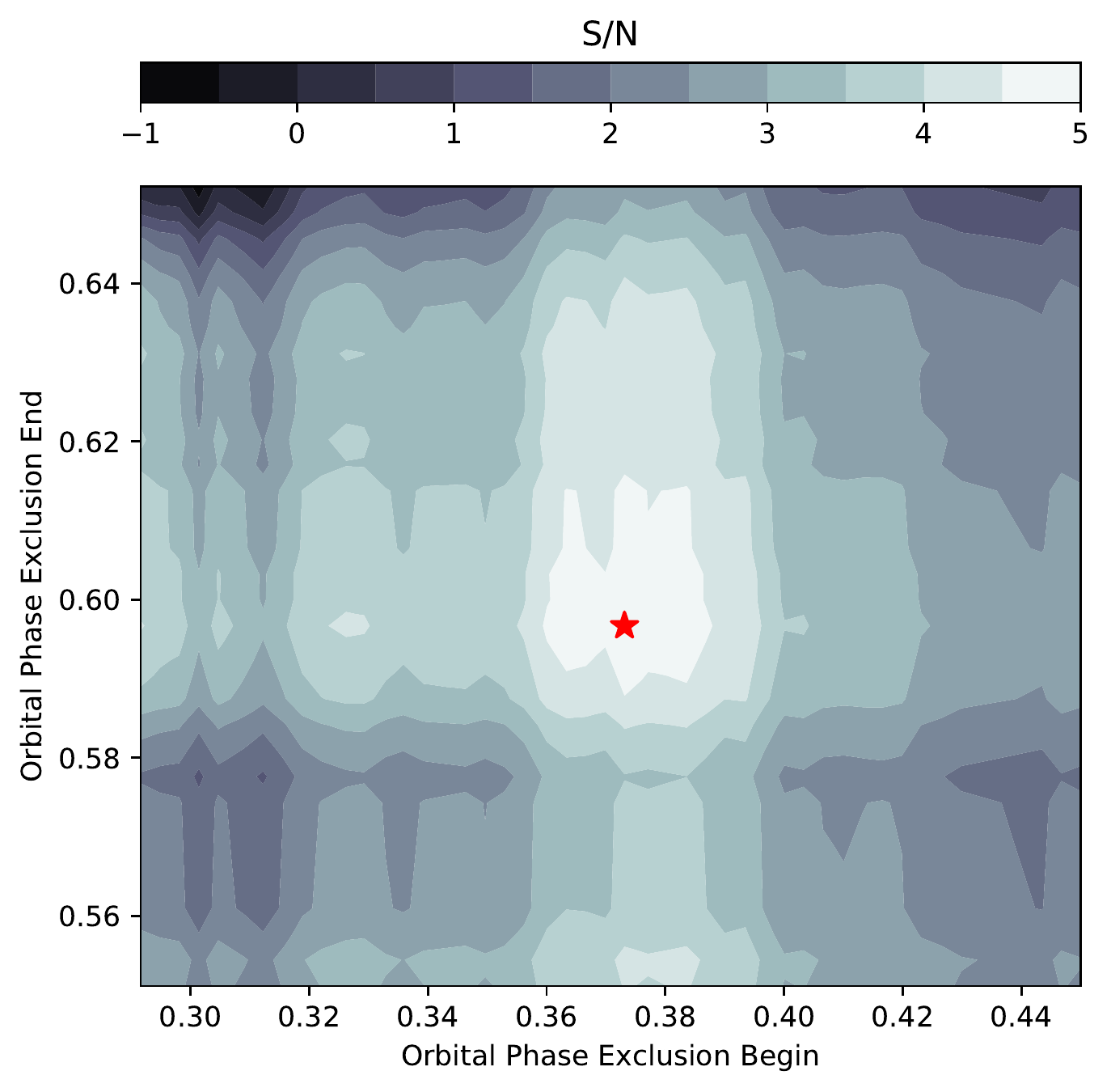}
	\caption{Map of TiO S/N detection significance when excluding different orbital phase ranges around the secondary eclipse (measured at $K_\mathrm{p}$\,=\,$248.0$\,km\,s$^{-1}$ and $\Delta \varv$\,=\,$0.0$\,km\,s$^{-1}$). The values on the x- and y-axis show the boundaries of the excluded phase intervals. The signal of TiO peaks when the orbital phase interval between $\sim$\,0.37 and $\sim$\,0.60 is excluded. The corresponding phase values are indicated with a red star (\textcolor{red}{$\star$}). We checked the exclusion of all possible orbital phase intervals. Phase ranges that are entirely inside eclipse were not considered (between roughly 0.45 and 0.55).}
	\label{phase-exclusion-mapping_TiO}
\end{figure*}

\begin{figure*}
	\centering
	\includegraphics[width=\textwidth]{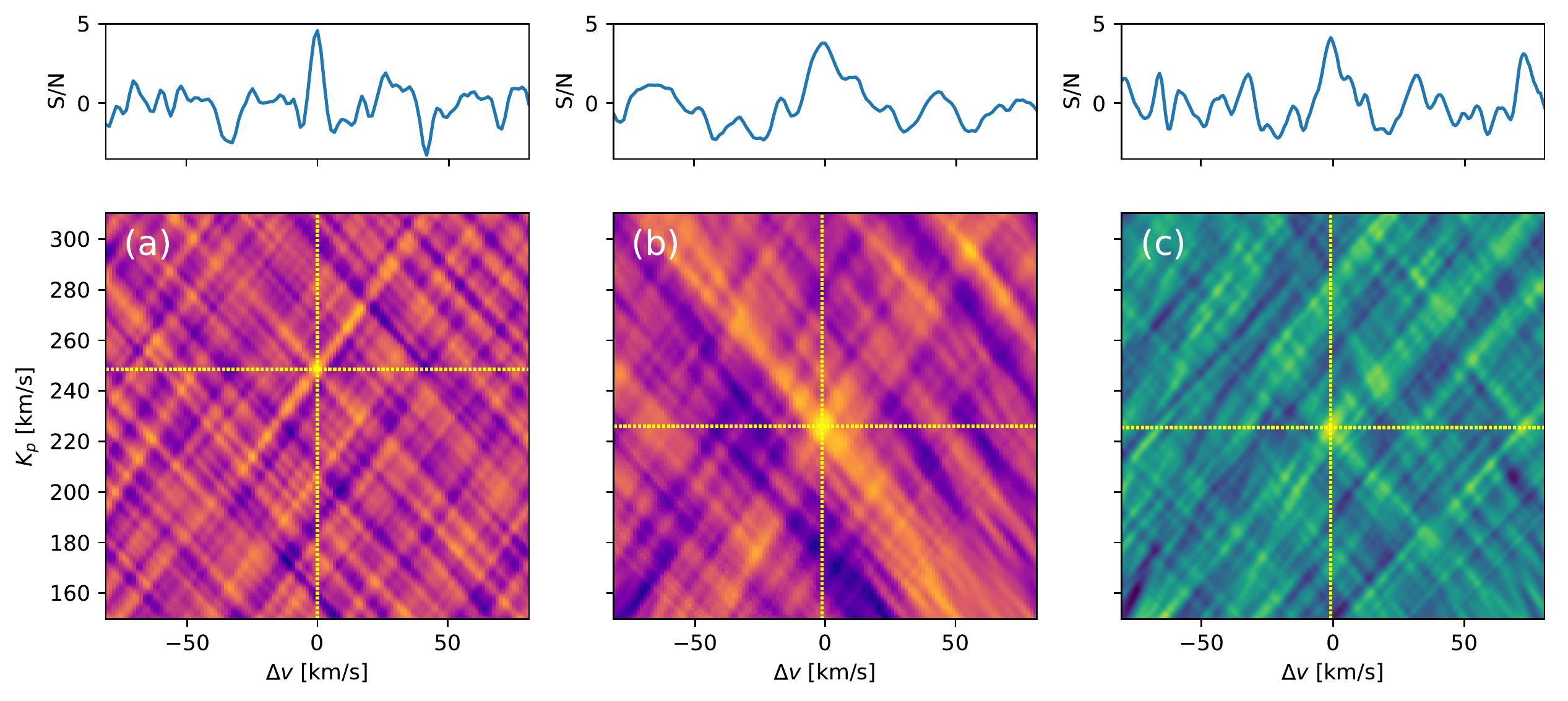}
	\caption{S/N maps obtained by exchanging the $T$-$p$ profiles of TiO and Fe to compute the model spectra for cross-correlation. All detection strengths are lower than those found in Sections \ref{Detection of Fe} and \ref{Evidence for TiO}. Panel \textit{(a)} corresponds to the TiO signal observed with CARMENES (S/N\,=\,4.6); panel \textit{(b)} is the Fe detection with CARMENES (S/N\,=\,3.8); panel \textit{(c)} is the Fe detection with HARPS-N (S/N\,=\,4.1). The horizontal panels show the cross-sections of the S/N peaks. Each S/N map corresponds to the \texttt{SYSREM} iteration number that maximizes the detection strength in Sections \ref{Detection of Fe} and~\ref{Evidence for TiO}.}
	\label{SN_maps_switch_Nugroho-vs-W189}
\end{figure*}

\begin{figure*}
	\centering
	\includegraphics[width=\textwidth]{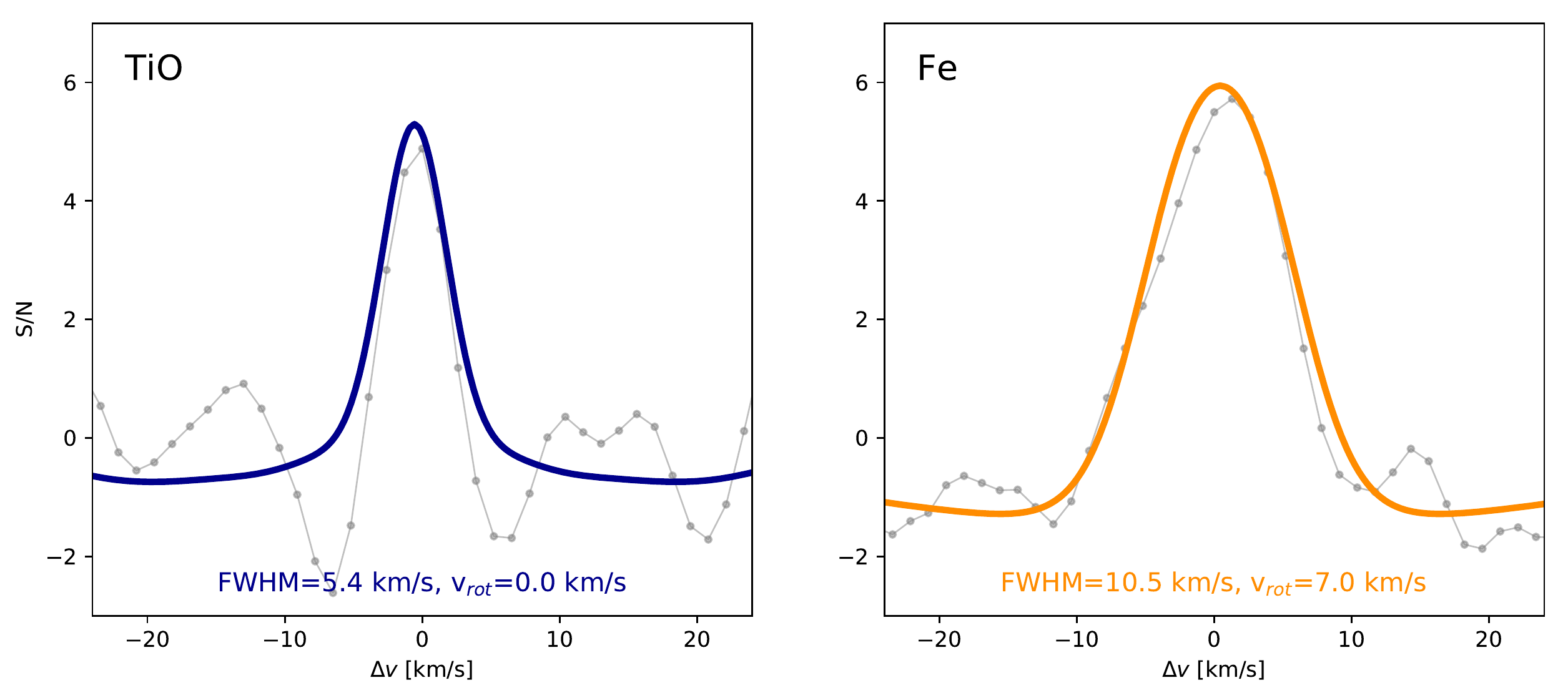}
	\caption{
		Auto-correlation of  the  TiO  model  spectrum (\textit{left} panel) and cross-correlation  between the Fe model spectrum and a rotationally broadened version of itself ($\varv_\mathrm{rot}$\,=\,7\,km\,s$^{-1}$; \textit{right} panel). For comparison, we plot the observed CCFs from the CARMENES observations in gray lines. The auto-correlation of the TiO model spectrum has a width of 5.4\,km\,s$^{-1}$. This is close to the FWHM value of the observed CCF ($4.0\pm0.7$\,km\,s$^{-1}$). Also the FWHM of the cross-correlation with the broadened Fe model spectrum is close to the value of the observed CCF (10.5\,km\,s$^{-1}$ and $8.6\pm1.0$\,km\,s$^{-1}$, respectively).}
	\label{autocorrelation-with-broadened-template}
\end{figure*}

\begin{figure*}
	\centering
	\includegraphics[width=\textwidth]{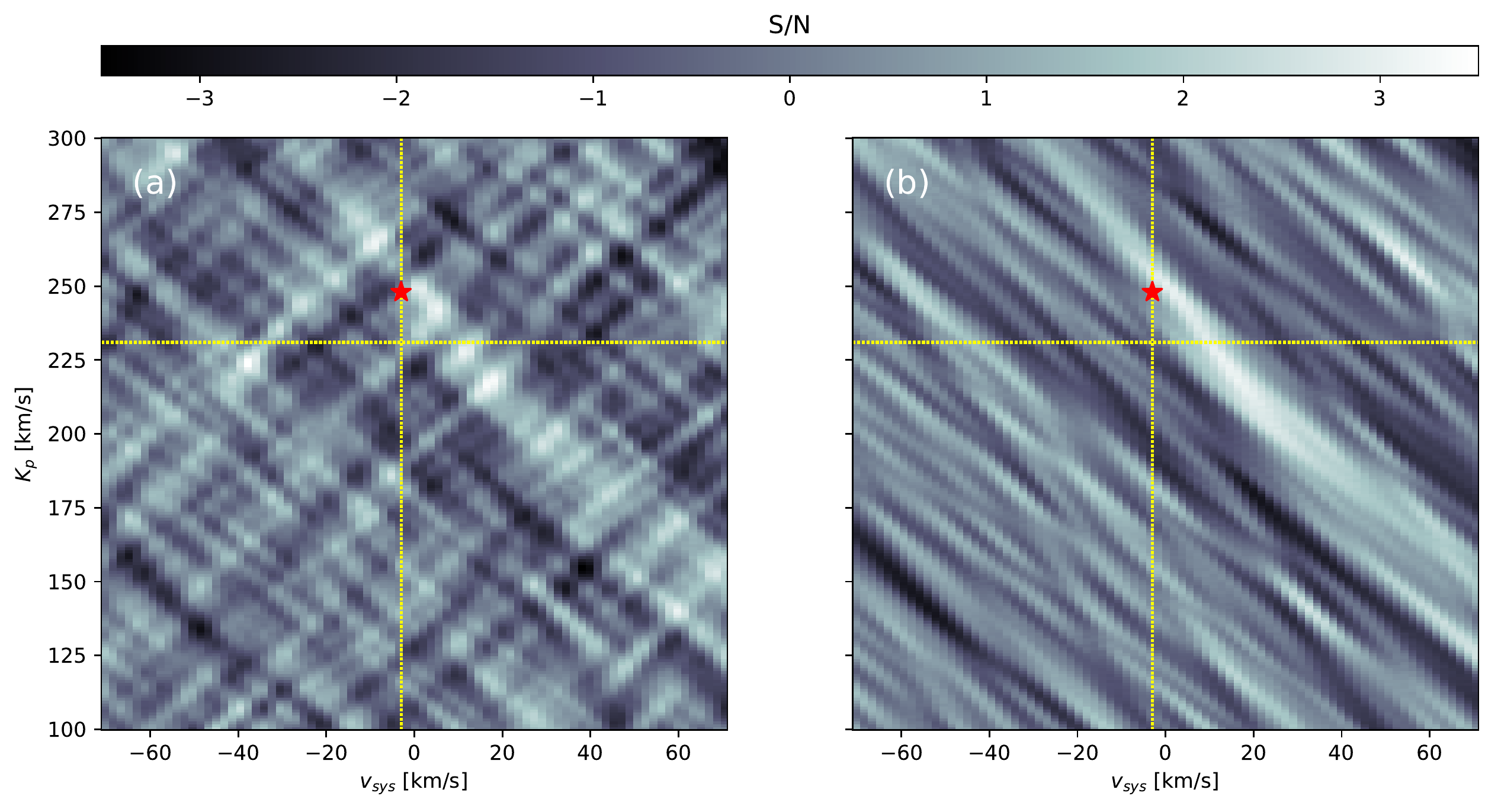}
	\caption{S/N maps of TiO from the ESPaDOnS observations. Panel \textit{(a)} represents the S/N map of all spectra; panel \textit{(b)} is the S/N map of the pre-eclipse TiO signature. The expected orbital parameters are indicated by the yellow dashed lines. A weak TiO signal is located close to the orbital parameters found with CARMENES, which is indicated with a red star (\textcolor{red}{$\star$}). Only the spectra before eclipse contribute to the signal. The x-axis is presented in the systemic rest frame ($\varv_{\mathrm{sys}}$) in order to be consistent with \cite{Herman2020}.}
	\label{SN_map_TiO_prior-to-eclipse_ESPADONS}
\end{figure*}

\end{document}